\pdfoutput=1

\documentclass[11pt,twoside,a4paper,cmspaper,final,collab]{cms-tdr}
\def\svnVersion{415899}\def\cmsCernNoTag{CERN-EP/2016-324}\def\cmsCernDate{\today}\def\cmsMessage{Published in the Journal of High Energy Physics as \href{http://dx.doi.org/10.1007/JHEP07(2017)003}{\doi{10.1007/JHEP07(2017)003.}}}

\begin{document}\cmsNoteHeader{TOP-12-039}

\hyphenation{had-ron-i-za-tion}
\hyphenation{cal-or-i-me-ter}
\hyphenation{de-vices}

\RCS$HeadURL: svn+ssh://svn.cern.ch/reps/tdr2/papers/TOP-12-039/trunk/TOP-12-039.tex $
\RCS$Id: TOP-12-039.tex 415811 2017-07-11 21:29:01Z cmackay $

\newcommand{\mwt}{\ensuremath{m_{\mathrm{T}}^{\mathrm{W}}}}
\newcommand\T{\rule{0pt}{2.6ex}}
\newcommand\B{\rule[-1.2ex]{0pt}{0pt}}
\newcommand{\eee}{\ensuremath{\Pe\Pe\Pe}}
\newcommand{\eemu}{\ensuremath{\Pe\Pe\mu}}
\newcommand{\mumue}{\ensuremath{\mu\mu\Pe}}
\newcommand{\mumumu}{\ensuremath{\mu\mu\mu}}

\newlength\cmsFigWidth
\ifthenelse{\boolean{cms@external}}{\setlength\cmsFigWidth{0.85\columnwidth}}{\setlength\cmsFigWidth{0.4\textwidth}}
\ifthenelse{\boolean{cms@external}}{\providecommand{\cmsLeft}{top\xspace}}{\providecommand{\cmsLeft}{left\xspace}}
\ifthenelse{\boolean{cms@external}}{\providecommand{\cmsRight}{bottom\xspace}}{\providecommand{\cmsRight}{right\xspace}}
\cmsNoteHeader{TOP-12-039}
\title{Search for associated production of a \PZ boson with a single top quark and for $\PQt \PZ$ flavour-changing interactions in $ \Pp \Pp $ collisions at $\sqrt{s} = 8\TeV $}

\date{\today}
\abstract{
A search for the production of a single top quark in association with a \PZ boson is presented, both to identify the expected standard model process and to search for flavour-changing neutral current interactions. The data sample corresponds to an integrated luminosity of 19.7\fbinv recorded by the CMS experiment at the LHC in proton-proton collisions at $\sqrt{s} = 8\TeV$. Final states with three leptons (electrons or muons) and at least one jet are investigated. An events yield compatible with $ \PQt \PZ \PQq $ standard model production is observed, and the corresponding cross section is measured to be $\sigma(\Pp\Pp \to \PQt \PZ \PQq \to \ell \nu \PQb \ell^+ \ell^- \PQq) = 10^{+8}_{-7}\unit{fb}$ with a significance of 2.4 standard deviations. No presence of flavour-changing neutral current production of $ \PQt \PZ \PQq $ is observed. Exclusion limits at 95\% confidence level on the branching fractions of a top quark decaying to a \PZ boson and an up or a charm quark are found to be $  { \cal{B} }( \PQt \rightarrow \PZ \PQu) < 0.022\%$ and $ { \cal{B} }( \PQt \rightarrow  \PZ \PQc ) < 0.049\%$. 
}
\hypersetup{%
pdfauthor={CMS Collaboration},%
pdftitle={Search for associated production of a Z boson with a single top quark and for tZ flavour-changing interactions in pp collisions at sqrt(s) = 8 TeV },%
pdfsubject={CMS},%
pdfkeywords={CMS, physics, software, computing}}
\maketitle
\section{Introduction}
\label{sec:Intro}
\begin{figure*}[hbtp]
\centering
{\includegraphics[width=0.35\textwidth]{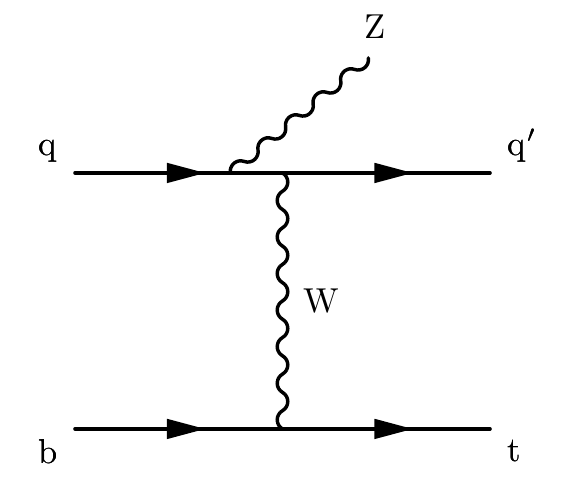}} 
{\includegraphics[width=0.35\textwidth]{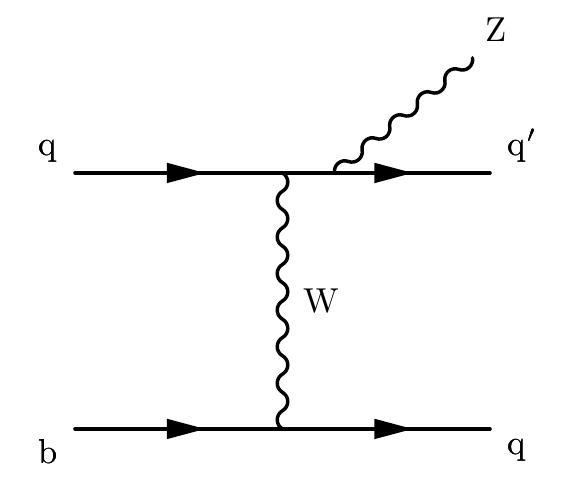}}\\
{\includegraphics[width=0.35\textwidth]{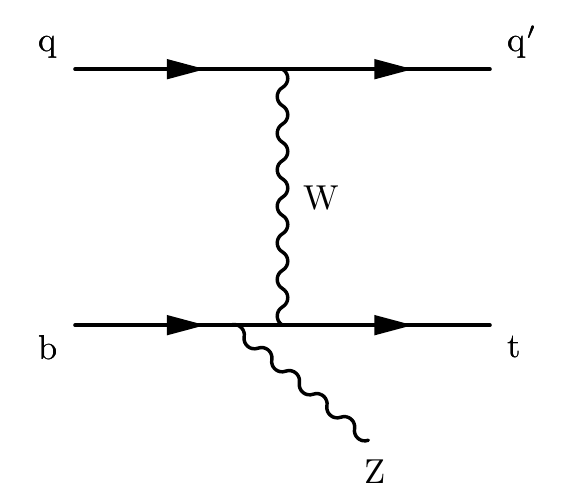}}
{\includegraphics[width=0.35\textwidth]{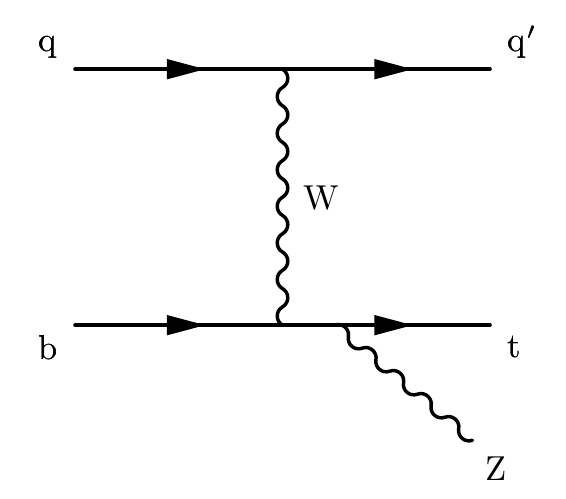}}\\
{\includegraphics[width=0.35\textwidth]{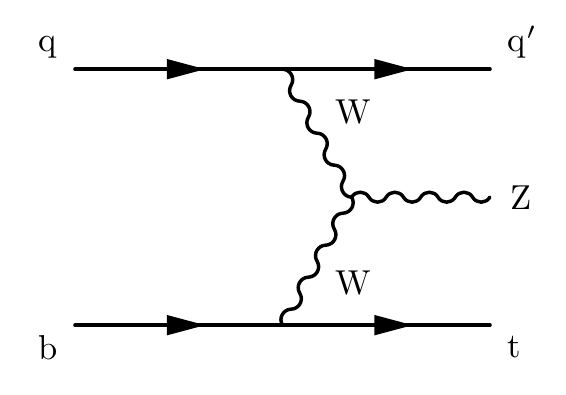}}
{\includegraphics[width=0.35\textwidth]{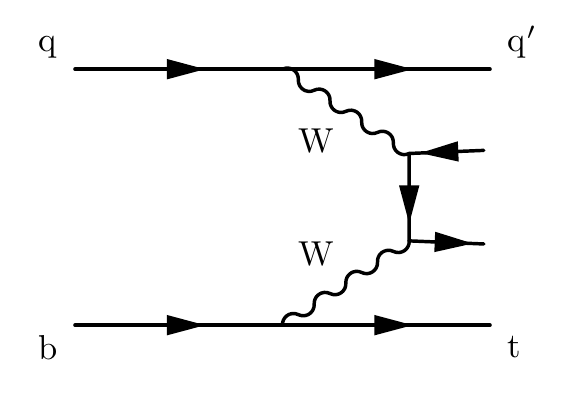}}\\
\caption{Leading-order $ \PQt \PZ \PQq $ production Feynman diagrams (all but bottom-right). The initial- and final-state quarks denoted $\PQq$ and $\PQq^\prime$ are predominantly first generation quarks, although there are smaller additional contributions from strange- and charm-initiated diagrams.  The bottom-right diagram represents the NLO nonresonant contribution to the $ \PQt \PZ \PQq $ process.}
\label{fig:feyn}
\end{figure*}

\begin{figure*}[hbtp]
\centering
{\includegraphics[width=0.4\textwidth]{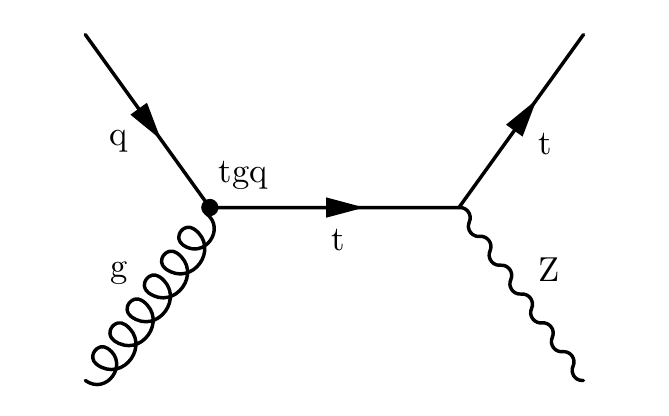}} 
{\includegraphics[width=0.4\textwidth]{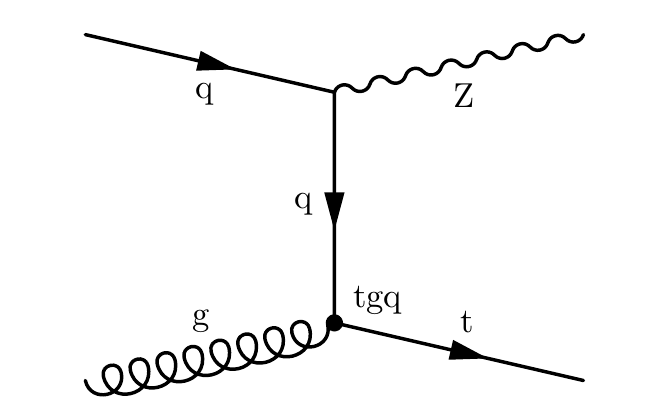}}\\
{\includegraphics[width=0.4\textwidth]{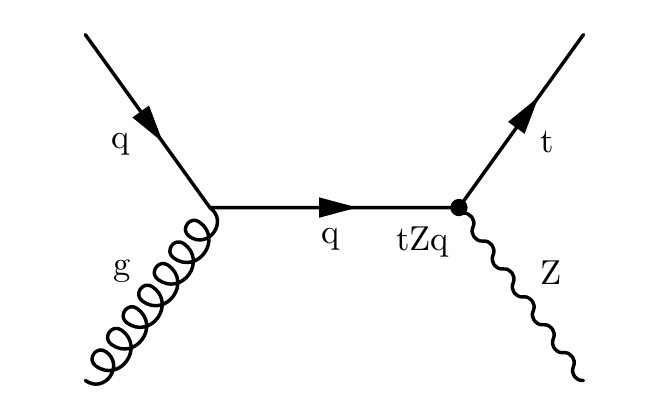}}
{\includegraphics[width=0.4\textwidth]{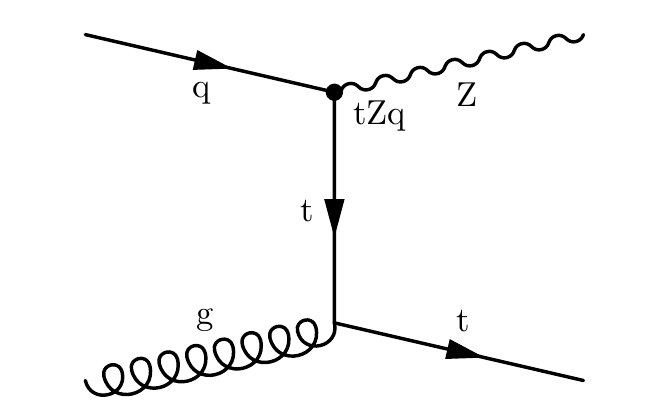}}\\
\caption{Feynman diagrams for the production of tZ in tZ-FCNC channels.}
\label{fig:feyn_FCNC}
\end{figure*}

\begin{figure*}[hbtp]
\centering
\resizebox{9cm}{!}{\includegraphics{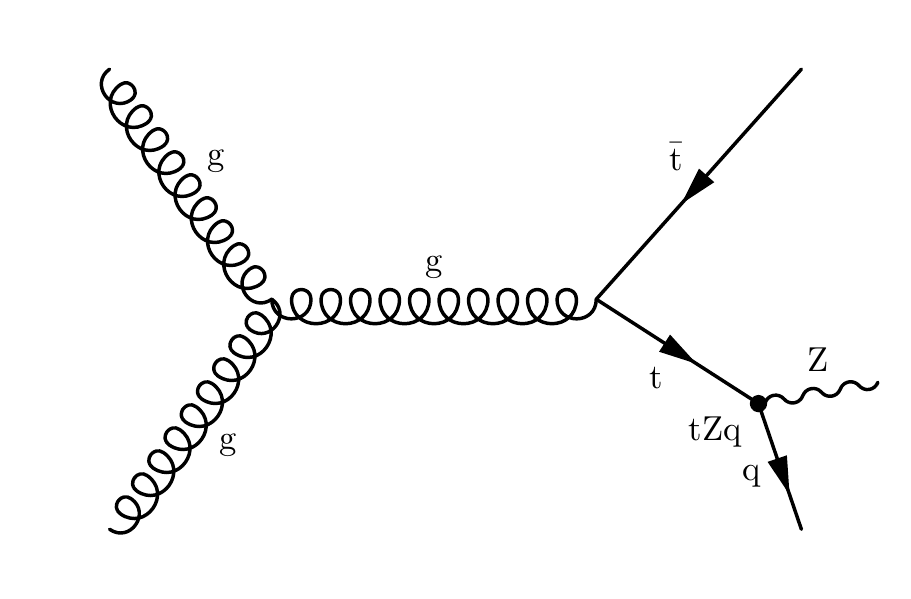}} 
\caption{Feynman diagram for the production of $ \PQt \PZ \PQq $ in the \ttbar-FCNC channel.}
\label{fig:feyn_FCNC_ttbar}
\end{figure*}

The top quark is the most massive particle in the standard model (SM) of particle physics.  Since its discovery in 1995~\cite{Abe:1995hr,Abachi:1995iq}, considerable advances have been made in understanding its properties.
At hadron colliders top quarks arise predominantly from the production of top quark-antiquark (\ttbar) pairs through the strong interaction. However, top quarks may also be produced singly from electroweak processes through three different production mechanisms. These are categorised by the virtuality of the W boson involved in the interaction: $t$-channel, $s$-channel and associated tW production. At the CERN LHC, the $t$- and tW channel production have been observed by the ATLAS and CMS Collaborations and their cross sections have been measured at both 7 and 8\TeV, respectively \cite{CMSPRL107,CMStchan7TeV,CMStchan8TeV,ATLAStchan7TeV,CMStWchan8TeV,ATLAStWchan8TeV}. The ATLAS and CMS Collaborations have recently published results of searches for $s$-channel single top quark production using 8\TeV data \cite{ATLASsChan,CMSschan}.
The high integrated luminosity and centre-of-mass energy at the LHC motivate the search for rare SM single top quark production processes, such as the production of a single top quark in association with a Z boson, where the top quark is produced via the $t$ channel and the Z boson is either radiated off one of the participating quarks or produced via W boson fusion (Fig.~\ref{fig:feyn}). These production mechanisms, referred to here as $ \PQt \PZ \PQq $-SM production, lead to a signature with a single top quark, a Z boson, and an additional quark. The process is sensitive to the coupling of the top quark to the Z boson, as illustrated in Fig.~\ref{fig:feyn}(middle-right). It is also related to WZ boson production, as can be seen in Fig.~\ref{fig:feyn}(bottom-left).
Thus, the observation of $ \PQt \PZ \PQq $ production and the subsequent measurement of the production cross section represent a test of the SM. 
The predicted $ \PQt \PZ \PQq $-SM production cross section for proton-proton collisions at a centre-of-mass energy of 8\TeV, at next-to-leading order (NLO), is \mbox{$\sigma(\Pp\Pp\to \rm \PQt \PZ \PQq ) = 236^{+11}_{-4} \text{ (scale)} \pm 11 \text{ (PDF)}\unit{fb}$} ~\cite{Campbell:2013yla}, where t denotes either a top quark or antiquark. The first uncertainty is associated with the renormalisation and factorisation scales used, and the second one is associated with the choice of parton distribution functions (PDFs). The CTEQ6M set of PDFs~\cite{Pumplin:2002vw} is used to determine the predicted cross section.
The cross section of the three-lepton final state, $\sigma(\Pp\Pp\to \PQt \ell^+\ell^- \PQq) \, { \cal{B} }(\PQt \to \ell \nu \PQb)$, where $\ell$ denotes a charged lepton (electron, muon, or tau), is calculated to be
\begin{center}
$\sigma(\Pp\Pp\to \PQt \ell^+\ell^- \PQq) \, { \cal{B} }(\PQt \to \ell \nu \PQb) = 8.2\unit{fb}$
\end{center} 
with a theoretical uncertainty of less than 10\%. The calculation is made in the five-flavour scheme, where b quarks are considered as coming from the interacting protons, with \MADGRAPH{}5\_a\MCATNLO~\cite{MCatNLO}, using the NNPDF (version 2) PDF set~\cite{nnpdf}. This includes lepton pairs from off-shell Z bosons with an invariant mass ${m_{\ell^+\ell^-} }> 50\GeV$. This cross section is used as a reference in this paper.
The ATLAS and CMS Collaborations have published results searching for $\ttbar \PZ$ production, which is also sensitive to the coupling of the top quark to the Z boson~\cite{Aad:2015eua,Khachatryan:2014ewa,TOP-12-036,Aaboud:2016xve}.  A production cross section of $\sigma(\Pp\Pp\to { \ttbar \PZ}) = 200^{+80}_{-70} \stat ^{+40}_{-30} \syst \unit{fb} $ was measured by CMS at 8\TeV~\cite{Khachatryan:2014ewa}. 
Within the SM, any flavour-changing neutral current (FCNC) involving the top quark and the Z boson, referred to here as tZ-FCNC, is forbidden at tree level and is suppressed at higher orders because of the GIM mechanism~\cite{GIM}. Some SM extensions, such as R-parity violating supersymmetric models~\cite{Yang:1997dk}, top-colour assisted technicolour models~\cite{Lu:2003yr} and singlet quark models~\cite{AguilarSaavedra:2002kr}, predict enhancements of the FCNC branching fraction, which could be as large as $\mathcal{O}(10^{-4})$~\cite{AguilarSaavedra:2004wm}. The production of a single top quark in association with a Z boson is sensitive to both $ \PQt \PZ \PQq $ and $ \PQt \Pg \PQq $ anomalous couplings~\cite{delAguila:1999ac,AguilarSaavedra:2004wm,Agram:2013koa} as shown in Figs.~\ref{fig:feyn_FCNC} and \ref{fig:feyn_FCNC_ttbar}. 
Searches for FCNC in the top quark sector have already been performed at the Fermilab Tevatron~\cite{CDFFCNC, DOFCNC} and at the LHC. The ATLAS Collaboration performed searches for anomalous $ \PQt \Pg \PQq $ couplings~\cite{ATLASarXiv:1509.00294} and the CMS Collaboration performed searches for $\PQt \gamma \PQq$ anomalous couplings~\cite{CMSarXiv:1511.03951}, while both the ATLAS and CMS Collaborations performed searches for $ \PQt \PZ \PQq $ anomalous couplings~\cite{ATLASarXiv:1508.05796, CMSTOPFCNC}. The most stringent exclusion limit at 95\% confidence level (CL) on the branching fraction ${ \cal{B} }( \PQt \rightarrow \PZ \PQq )$,  set by the CMS Collaboration, excludes branching fractions greater than 0.05\% \cite{CMSTOPFCNC}.
In this paper, two separate searches, using similar event selections and background estimates, are presented: a search for $ \PQt \PZ \PQq $-SM production and a search for tZ-FCNC production from anomalous couplings. Both searches are performed using a data set of proton-proton collisions  at a centre-of-mass energy of 8\TeV, corresponding to an integrated luminosity of 19.7\fbinv. In $ \PQt \PZ \PQq $-SM production, because the processes involved are based on $t$-channel single top quark production, the signature consists of a single top quark, a Z boson, and an additional jet preferentially emitted in the forward region of the detector (absolute pseudorapidity $|\eta|> 2.4$). The search for tZ-FCNC is performed by combining the single top quark and \ttbar\ production modes. The single top quark production leads to a signature containing a top quark and a Z boson (single-top-quark-FCNC) with no extra jets from the matrix-element calculation. For the \ttbar\ production mode (\ttbar-FCNC), the FCNC vertex appears in the decay of the top quark, and leads to the same signature as for $ \PQt \PZ \PQq $-SM, but with the jet not associated with the b quark being produced in the central region of the detector. Both searches are performed in the trilepton final state, where both the W boson from the top quark and the Z boson decay into either electrons or muons, resulting in four possible leptonic combinations in the final state:  \eee, \mumumu, \mumue, and \eemu.  As they are not specifically excluded, there is also a contribution from leptonic $\tau$ decays. The main sources of background to these searches are \ttbar production, single top quark production, diboson production, $\ttbar \mathrm{V}$ (V = W or Z) and Drell--Yan (DY) production. The $ \PQt \PZ \PQq $-SM production is a key irreducible background to the FCNC search. The discrimination between signal and background is achieved using a boosted decision tree (BDT) and the nonprompt backgrounds are estimated from the data, whereas other backgrounds are estimated from simulation using constraints from data.
\section{Theoretical framework}
\label{sec:theo}
The generation of the $ \PQt \PZ \PQq $-SM events is performed at NLO using the \MADGRAPH{}5\_a\MCATNLO v5.1.3.30 generator~\cite{MCatNLO}. For the tZ-FCNC production, the description and generation of signal events follow the strategy detailed in Ref.~\cite{Agram:2013koa}. The generation is achieved by describing the relevant interactions in terms of a set of effective operators that are independent of the underlying theory. The searches are thus performed in a model-independent way.
The signature corresponding to the tZ-FCNC processes can be produced both via strong $ \PQt \Pg \PQq $ and weak $ \PQt \PZ \PQq $ couplings, as illustrated in Fig.~\ref{fig:feyn_FCNC}.  The \ttbar-FCNC production, where the anomalous coupling appears in the top quark decay, is presented in Fig.~\ref{fig:feyn_FCNC_ttbar}. Both of these production modes can be incorporated into the SM Lagrangian $\cal{L}$ using effective operators of dimensions 4 and 5~\cite{Agram:2013koa}:
\begin{align}
\begin{split}
\Lumi &= \sum_{\rm q=u,c} \Bigl[ \sqrt{2}g_{s} \frac{\kappa_{\rm tgq}}{\Lambda} {\rm \bar{t}} \sigma^{\mu\nu}T_{\rm a} ( f_{\rm q}^{\rm L}P_{\rm L} + f_{\rm q}^{\rm R}P_{\rm R} ){\rm q} G_{\mu\nu}^{\rm a} \\
&+ \frac{g}{\sqrt{2}c_{\rm W}} \frac{\kappa_{\mathrm{tZq}}}{\Lambda} {\rm \bar{t}} \sigma^{\mu\nu} ( \hat{f}_{\rm q}^{\rm L}P_{\rm L} + \hat{f}_{\rm q}^{\rm R}P_{\rm R} ) {\rm q Z}_{\mu\nu} \label{equ:FCNClag} \\
&+ \frac{g}{4c_{\rm W}} \zeta_{\mathrm{tZq}} {\rm \bar{t}} \gamma^{\mu}  ( \bar{f}_{\rm q}^{\rm L}P_{\rm L} + \bar{f}_{\rm q}^{\rm R}P_{\rm R} ) {\rm q Z}_{\mu} \Bigr] \text{+ h.c.}
\end{split}
\end{align}
The effects of new physics contributions are quantified through the dimensionless parameters $\kappa_{\rm tgq}$, $\kappa_{\mathrm{tZq}}$, and $\zeta_{\mathrm{tZq}}$ together with the complex chiral parameters $f\,_{\rm q}^{\rm L,R}$, $\hat{f}\,^{\rm L,R}_{\rm q}$, and $\bar{f}\,^{\rm L,R}_{\rm q}$, which can be constrained as $\vert f_{\rm q}^{\rm L} \vert^{2} + \vert f_{\rm q}^{\rm R} \vert^{2} = \vert \hat{f}_{\rm q}^{\rm L} \vert^{2} + \vert \hat{f}_{\rm q}^{\rm R} \vert^{2} = \vert \bar{f}_{\rm q}^{\rm L} \vert^{2} + \vert \bar{f}_{\rm q}^{\rm R} \vert^{2} = 1$. The energy scale at which these effects are assumed to be relevant is parametrised by $\Lambda$.
The two couplings to the gluon, ${\kappa_{\rm tgu}}/{\Lambda}$ and ${\kappa_{\rm tgc}}/{\Lambda}$, relate to the diagrams shown at the top of  Fig.~\ref{fig:feyn_FCNC}, while the four couplings to the Z boson, ${\kappa_{\rm tZu}}/{\Lambda}$, $\zeta_{\rm tZu}$, ${\kappa_{\rm tZc}}/{\Lambda}$, and $\zeta_{\rm tZc}$  relate to the diagrams shown at the bottom of Fig.~\ref{fig:feyn_FCNC}. The anomalous couplings related to the weak and strong sectors are assumed to be independent of each other, although interference is expected to occur between the ${\kappa_{\mathrm{tZq}}}/{\Lambda}$ and $\zeta_{\mathrm{tZq}}$ contributions. The sensitivity to the ${\kappa_{\rm tgq}}/{\Lambda}$ coupling is poor in comparison to other channels \cite{ATLASarXiv:1509.00294}, while $\zeta_{\mathrm{tZq}}$ couplings lead to very small cross sections \cite{Agram:2013koa}. For these reasons we consider here only cases where ${\kappa_{\mathrm{tZq}}}/{\Lambda} \neq 0$, while setting $\zeta_{\mathrm{tZq}}=0$  and ${\kappa_{\rm tgq}}/{\Lambda}=0$. Furthermore, the interference between single top quark and \ttbar-FCNC processes is neglected and the 4 fermion interactions are not included in this analysis~\cite{Durieux:2014xla}.
\section{CMS detector}
The central feature of the CMS apparatus is a superconducting solenoid of 6\unit{m} internal diameter, providing a magnetic field of 3.8\unit{T}. Within the solenoid volume are a silicon pixel and strip tracker, a lead tungstate crystal electromagnetic calorimeter (ECAL), and a brass and scintillator hadron calorimeter (HCAL), each composed of a barrel and two endcap sections. Forward calorimeters extend the pseudorapidity coverage provided by the barrel and endcap detectors. Muons are measured in gas-ionisation detectors embedded in the steel flux-return yoke outside the solenoid.
The ECAL provides coverage in pseudorapidity $\abs{ \eta }< 1.48 $ in the barrel region and $1.48 <\abs{ \eta } < 3.0$ in two endcap regions (EE). A preshower detector consisting of two planes of silicon sensors interleaved with a total of $3 X_0$ of lead is located in front of the EE. The electron momenta are estimated by combining energy measurements in the ECAL with momentum measurements in the tracker~\cite{Khachatryan:2015hwa}. The relative transverse momentum resolution for electrons with $\pt {\approx} 45\GeV$ from $\Z \rightarrow \Pe \Pe$ decays ranges from 1.7\% in the barrel region to 4.5\% in the endcaps~\cite{Khachatryan:2015hwa}. The dielectron mass resolution for $\Z \rightarrow \Pe \Pe$ decays when both electrons are in the ECAL barrel is 1.9\%, and is 2.9\% when both electrons are in the endcaps. 
Muons are measured in the range $\abs{\eta}< 2.4$. Matching muons to tracks measured in the silicon tracker results in a relative \pt resolution for muons with $20 <\pt < 100\GeV$ of 1.3--2.0\% in the barrel and better than 6\% in the endcaps. The \pt resolution in the barrel is better than 10\% for muons with \pt up to 1\TeV~\cite{Chatrchyan:2012xi, Muon}. 
Events of interest are selected using a two-tiered trigger system~\cite{Khachatryan:2016bia}. The first level, composed of custom hardware processors, uses information from the calorimeters and muon detectors to select events at a rate of around 100\unit{kHz} within a time interval of less than 4\mus. The second level, known as the high-level trigger, consists of a farm of processors running a version of the full event reconstruction software optimised for fast processing, and reduces the event rate to less than 1\unit{kHz} before data storage. 
A more detailed description of the CMS detector, together with a definition of the coordinate system used and the relevant kinematic variables, can be found in
Ref.~\cite{Chatrchyan:2008zzk}. 
\section{Monte Carlo simulation}
\label{sec:pdf}
Simulated $ \PQt \PZ \PQq $-SM and $\ttbar\PZ$ events are produced, at NLO, with the \MADGRAPH{}5\_a\MCATNLO v5.1.3.30 generator~\cite{MCatNLO}, interfaced with {\sc pythia} version 8.212~\cite{pythia8} for parton showering and hadronisation.  Several of the background processes considered in this analysis ($ \ttbar $ and $ \ttbar\PW $ production, diboson production and Z boson production in association with multiple jets) are produced at leading order (LO) using the \MADGRAPH{}5\_a\MCATNLO Monte Carlo (MC) generator interfaced with {\sc pythia} version 6.426~\cite{Sjostrand:2006za}. Single top quark background processes (tW and $\rm {\bar{t}}$W) are simulated using the \POWHEG v.1.0 r1380 generator~\cite{Re:2010bp,Alioli:2010xd,Alioli:2009je,Frixione:2007vw}, which is interfaced to {\sc pythia} version 8.212 for parton showering and hadronisation. The tZ-FCNC events are generated at LO using the \MADGRAPH{}5\_a\MCATNLO generator interfaced with {\sc pythia} version 6.426. The $\kappa$ Lagrangian terms presented in Eq. (\ref{equ:FCNClag}) are implemented as a new model in \MADGRAPH{}5\_a\MCATNLO by means of the {\sc FeynRules} package \cite{Alloul:2013bka} and of the universal {\sc FeynRules} output format \cite{Degrande:2011ua}. The complex chiral parameters  are fixed to the following values:  $ \hat{f}_{\rm{q}}^{\rm {R}}  = 0$ and $\hat{f}_{\rm{q}}^{\rm{L}} = 1$.
All samples generated with \POWHEG\ and \MADGRAPH{}5\_a\MCATNLO use the CT10 \cite{CT10}  PDF set.  The value of the top quark mass used in all the simulated samples is $m_{\rm t} = 172.5\GeV$. All samples include W boson decays to $\tau$ leptons, as well as to electrons and/or muons. The  characterisation of the underlying event  uses the {\sc pythia}\  Z2* tune \cite{Field:2010bc,Field:2} for the \MADGRAPH{}5\_a\MCATNLO and \POWHEG\ samples, and the CUETP8M1 tune \cite{Field:2} for the $ \PQt \PZ \PQq $-SM sample.
Additional samples of $ \PQt \PZ \PQq $-SM, tZ-FCNC, $ \ttbar\mathrm{V} $, and WZ are generated, varying the renormalisation and factorisation scales, for studies of systematic effects. For the $ \ttbar\mathrm{V} $ and WZ backgrounds, further samples are generated varying the merging threshold in \MADGRAPH{}5\_a\MCATNLO. 
The expected cross sections are obtained from next-to-next-to-leading-order calculations for $\rm t\bar{t}$~\cite{Czakon:2011xx}  and  Z/$\gamma^*$ processes~\cite{Melnikov:2006kv}, NLO plus next-to-next-to-leading-logarithmic calculations for single top quark production in the tW or  $ \ttbar \PW $ channels \cite{Kidonakis:2010ux}, and NLO calculations for VV~\cite{Campbell:2011bn} and $ \ttbar\mathrm{V} $~\cite{Campbell:2012dh,Garzelli:2012bn} processes.
For all samples of simulated events, multiple minimum-bias events generated with {\sc pythia} are added to simulate the presence of additional proton-proton interactions (pileup) from the same bunch crossing or in neighbouring proton bunches. To refine the simulation, the events are weighted to reproduce the distribution in the number of pileup vertices inferred from data.  Most generated samples contain full simulation of detector effects, using the \GEANTfour package~\cite{Agostinelli:2002hh}, including simulation of the machine running conditions, while the FCNC samples are processed using a fast simulation of the detector ~\cite{fastsim}.
\section{Event reconstruction and data selection}
In the searches presented in this paper, the signal signature contains a Z boson and a top quark, which both decay leptonically to either electrons or muons. Thus the final state for both searches consists of three leptons (electrons and/or muons, including those coming from tau decays), plus an escaping undetected neutrino that is inferred from an imbalance in the transverse momentum. The signature also includes a bottom quark jet (b jet) that arises from the hadronisation of the b quark produced in the top quark decay. In the final state for $ \PQt \PZ \PQq $-SM production, or for \ttbar-FCNC, there is an additional jet arising from the hadronisation of a light or a charm quark. 
The data used in this analysis were collected with the CMS detector during the 2012 proton-proton data taking period at a centre-of-mass energy of 8\TeV. The data are selected online using triggers that rely on the presence of two high-\pt leptons, $\Pe \Pe$, $ \Pe \mu$, or $\mu\mu$.  The highest-\pt lepton is required to satisfy $\pt > 17\GeV$, while the second-highest-\pt lepton must satisfy $\pt > 8 \GeV$. In addition, the trigger selection requires loose lepton identification for both lepton flavours; electrons are additionally required to pass online isolation requirements. The resulting trigger efficiencies are 99\% for \eee\ and \eemu, 98\% for \mumumu\ and 89\% for \mumue. For tZ-FCNC production, the trigger acceptance is enhanced by using single-lepton and trilepton triggers with various \pt thresholds, resulting in a trigger efficiency close to 100\%, after all selection cuts. The trigger efficiency is obtained from data collected with an independent trigger selection based on missing transverse momentum. The missing transverse momentum vector \ptvecmiss is defined as the projection on the plane perpendicular to the beams of the negative vector sum of the momenta of all reconstructed particles in an event. Its magnitude is referred to as missing transverse momentum, \ptmiss.
A particle-flow event reconstruction algorithm~\cite{CMS-PAS-PFT-09-001,CMS-PAS-PFT-10-001} identifies each individual particle with an optimised combination of information from the various elements of the CMS detector. The energy of the photons is directly obtained from the ECAL measurement. The energy of the electrons is determined from a combination of the electron momentum at the primary interaction vertex as determined by the tracker, the energy of the corresponding ECAL cluster, and the energy sum of all bremsstrahlung photons spatially compatible with originating from the electron track. The momentum of the muons is obtained from the curvature of the corresponding track.  The energy of the charged hadrons is determined from a combination of their momentum measured in the tracker and the matching ECAL and HCAL energy deposits. Finally, the energy of the neutral hadrons is obtained from the corresponding corrected ECAL and HCAL deposits.
The tracks reconstructed in the silicon tracker are used to identify and construct a series of interaction vertices, which correspond to the pileup. For each vertex, the sum of the transverse momenta squared of the associated tracks is calculated. The vertex whose sum is largest is taken to be the event primary vertex, provided that it is reconstructed using four or more tracks and that it lies within $24\cm$ of the nominal interaction point in the $z$ direction and within $2\cm$ in the transverse plane.
Each event must contain exactly three electrons and/or muons, reconstructed by the particle-flow algorithm.  Each lepton must have $\pt > 20\GeV$ and $|\eta| < 2.5$ (electron) or $|\eta| < 2.4$ (muon) and must be isolated.  Isolation is determined by calculating the sum of \pt of all the other reconstructed particles that lie within a cone of fixed radius $\Delta R = \sqrt{\smash[b]{(\Delta\eta)^{2} + (\Delta\phi)^{2}}}$ around the lepton, correcting for the expected contribution from pileup~\cite{Perloff} and dividing the corrected sum by the $\pt$ of the lepton.  The resulting quantity is denoted $I_{ \rm rel}$. For electrons, the cone size is set to $\Delta R = 0.3$ and $I_{ \rm rel}$ must be less than $0.15$. For muons, the cone size is set to $\Delta R = 0.4$ and $I_{ \rm rel}$ must be less than $0.12$. Events that contain additional leptons, satisfying the same kinematic selection but with relaxed lepton identification criteria, are rejected. Lepton isolation and identification efficiencies in simulation are corrected to match the ones measured in data using a tag-and-probe method~\cite{Chatrchyan:2014mua}.
Two of the same-flavour leptons in each event are required to have opposite electric charge, and have an invariant mass, $m_{\ell \ell}$, compatible with the Z boson mass, i.e. \mbox{$76 < m_{\ell \ell} < 106 \GeV$}. In the \eee\ and \mumumu\ channels, the pair of oppositely charged leptons having an invariant mass closest to the Z boson mass is used to form the Z boson candidate. In the \eemu\ and \mumue\ channels, the same-flavour leptons are used to form the Z boson candidate. For all channels, the third lepton is assumed to come from the decay of the W boson. 
Jets are clustered from the particles reconstructed using the particle-flow algorithm with the infrared and collinear safe anti-\kt algorithm~\cite{Cacciari:2008gp, Cacciari:2011ma}, operated with a distance parameter $R$ = 0.5. Jet momentum is determined as the vectorial sum of all particle momenta in the jet, and is found from simulation to be within 5 to 10\% of the true particle-level jet momentum over the whole $\pt$ spectrum and detector acceptance. An offset correction is applied to jet energies to take into account the contribution from pileup interactions. Corrections for the jet energy are derived from simulation, and are corrected with in situ measurements of the energy balance in dijet and photon+jet events~\cite{Khachatryan:2016kdb}. For the tZ-FCNC analysis, only jets that satisfy $\pt > 30 \GeV$ and $|\eta| < 2.4$ are used in the results presented here, while for the $ \PQt \PZ \PQq $-SM analysis, the maximum allowed value of $|\eta|$ is relaxed to 4.5 to improve the signal acceptance, as for single top quark $t$-channel processes the extra light jet is mostly produced in the forward region. Jets that are reconstructed close to a selected lepton ($\Delta R < 0.5$) are removed.
Jets that originate from the hadronisation of a b quark are identified (tagged) using the combined secondary vertex algorithm~\cite{CMS-PAS-BTV-13-001}.  This algorithm combines various track-based variables with vertex-based variables to construct a discriminating observable in the region $|\eta| < 2.4$.
The discriminant is used to distinguish between b jets and non-b jets. For the results presented here, the so-called {\it loose} operating point is used. This corresponds to a b tagging efficiency of about 85\% and a misidentification probability of 10\% for light-flavour or gluon jets, as estimated from QCD multijet simulations. The value of the b tagging discriminant is also used in the multivariate discriminator. Corrections to the b tagging discriminant shape have been determined using $ \ttbar $ and multijet control samples, and are then applied to the signal and background data sets~\cite{CMS-PAS-BTV-13-001}.
In the search for $ \PQt \PZ \PQq $-SM production, two or more selected jets are required, one or more of which must also satisfy the b tagging requirements. In the search for tZ-FCNC production, two different signal selections are considered. In a first selection, denoted as single-top-quark-FCNC selection, exactly one selected jet is required, which has to pass the b tagging requirement. A second selection (\ttbar-FCNC selection) asks for at least two selected jets with at least one passing the b tagging requirement. The selections result in a signal-enriched sample, with either single-top-quark-FCNC or \ttbar-FCNC events. 
To further reject backgrounds, two additional selections are made on the missing transverse momentum and the transverse mass of the W boson, \mwt. These selections are applied to the signal regions only and are optimised to maximise the expected significance. The optimisation is made for the $ \PQt \PZ \PQq $-SM and tZ-FCNC signals separately. For the $ \PQt \PZ \PQq $-SM analysis, \mbox{$\mwt>10\GeV$} is required while for the tZ-FCNC analysis we require $\ptmiss > 40\GeV$ and $\mwt >10\GeV$. These selections define the signal regions for the analyses.
In addition to the signal region, a background-enriched control region is defined by requiring one or two selected jets, but vetoing events containing a b-tagged jet, in order to increase the DY and WZ content. The event selections for the control and signal regions are presented in Table~\ref{tab:select}, while the number of events remaining for each process, after all selections have been applied is shown in Table~\ref{tab:eventselect} for the tZq-SM shape analysis.
\begin{table}[!ht]
\begin{center}
\topcaption{The event selections for the signal and control regions for the SM and FCNC analyses.} 
\begin{tabular}{c|c|c|c|c}
\hline
SM signal  & SM control & FCNC signal & FCNC signal & FCNC control \\
tZq    & WZ & single-top-quark & $\ttbar$ & WZ \\ \hline
$\geqslant$2 jets, $|\eta|\!\!<$4.5&  1 or 2 jets, $|\eta|\!\!<$4.5  & 1 jet, $|\eta|\!\!<$2.4 & $\geqslant$ 2 jets, $|\eta|\!\!<$2.4 & 1 or 2 jets, $|\eta|\!\!<$2.4 \\
$\geqslant$ 1 b tag       &  0 b tag   &  1 b tag & $\geqslant$ 1 b tag & 0 b tag \\
$\mwt$ $>$ 10 GeV    & & $\mwt$ $>$ 10 GeV    & $\mwt$ $>$ 10 GeV &  \\
                 &    &  $\ptmiss$ $>$ 40 GeV &  $\ptmiss$ $>$ 40 GeV &  \\
\hline
\end{tabular}
\label{tab:select}
\end{center}
\end{table} 
\begin{table}[!ht]
\begin{center}
\topcaption{The number of events remaining for each process, after all selections have been applied, in the control and signal regions for the tZq-SM shape analysis. WZ+h.f. denotes WZ + heavy flavour.} 
\begin{tabular}{l|c|c}
\hline
Process & Control Region & Signal Region \\
\hline
$\ttbar\PZ$      &   1.76$\pm$0.18   &  10.91$\pm$0.44 \\
ZZ               &  10.64$\pm$0.03   &   1.58$\pm$0.01 \\
WZ+h.f.          & 104.73$\pm$1.32   &  34.34$\pm$0.76 \\
WZ               & 426.92$\pm$2.67   &  58.00$\pm$0.98 \\
DY               & 192.95$\pm$13.89  &  49.24$\pm$7.02 \\
tZq              &   5.89$\pm$0.03   &  16.05$\pm$0.04 \\ \hline
Total prediction & 743   $\pm$18     & 170   $\pm$9    \\ 
Data             & 763               & 154             \\
\hline
\end{tabular}
\label{tab:eventselect}
\end{center}
\end{table}
\section{Analysis method}
\begin{figure}[htbp!]
\begin{center}
{\includegraphics[width=0.45\linewidth]{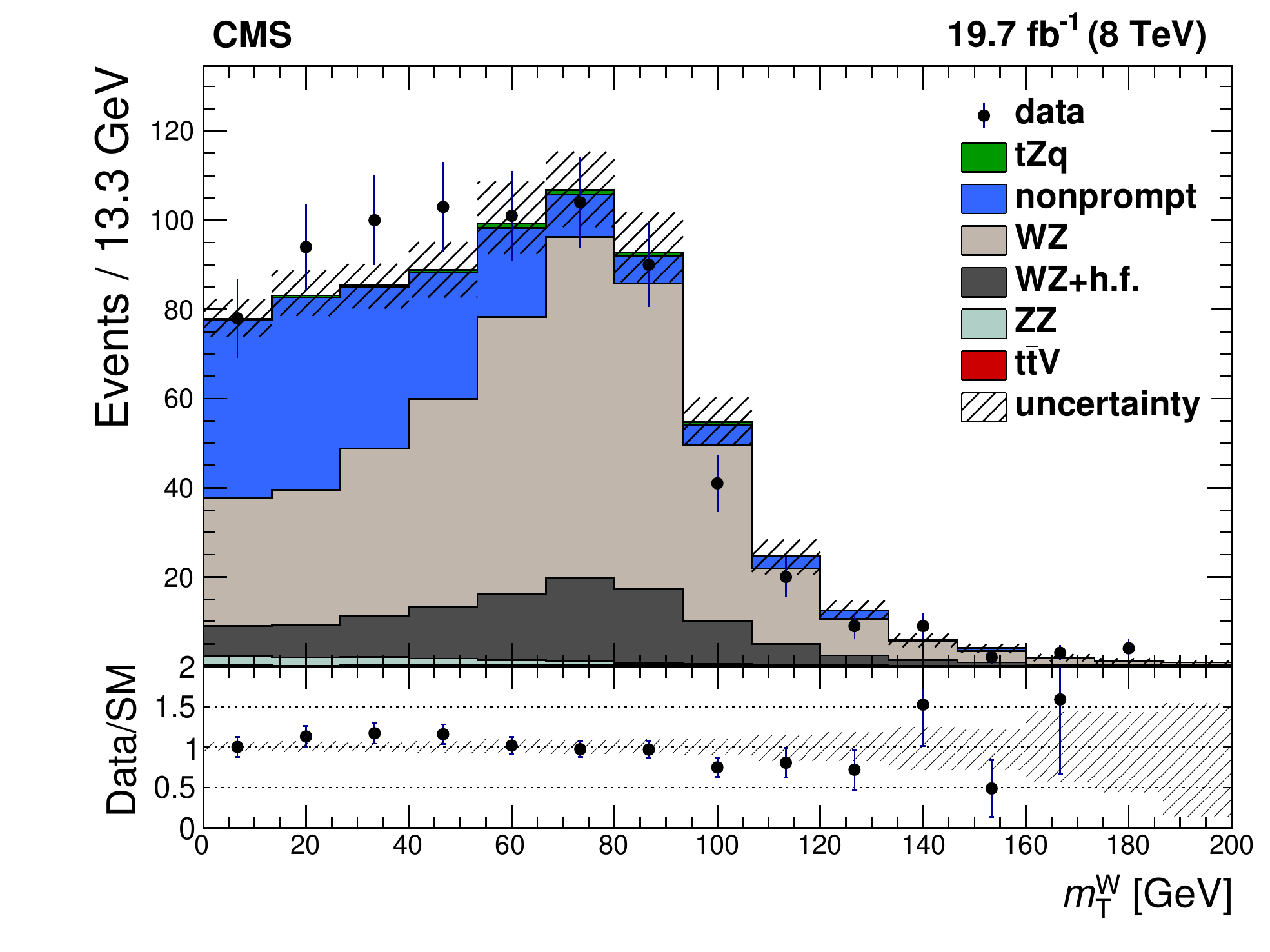}}
{\includegraphics[width=0.45\linewidth]{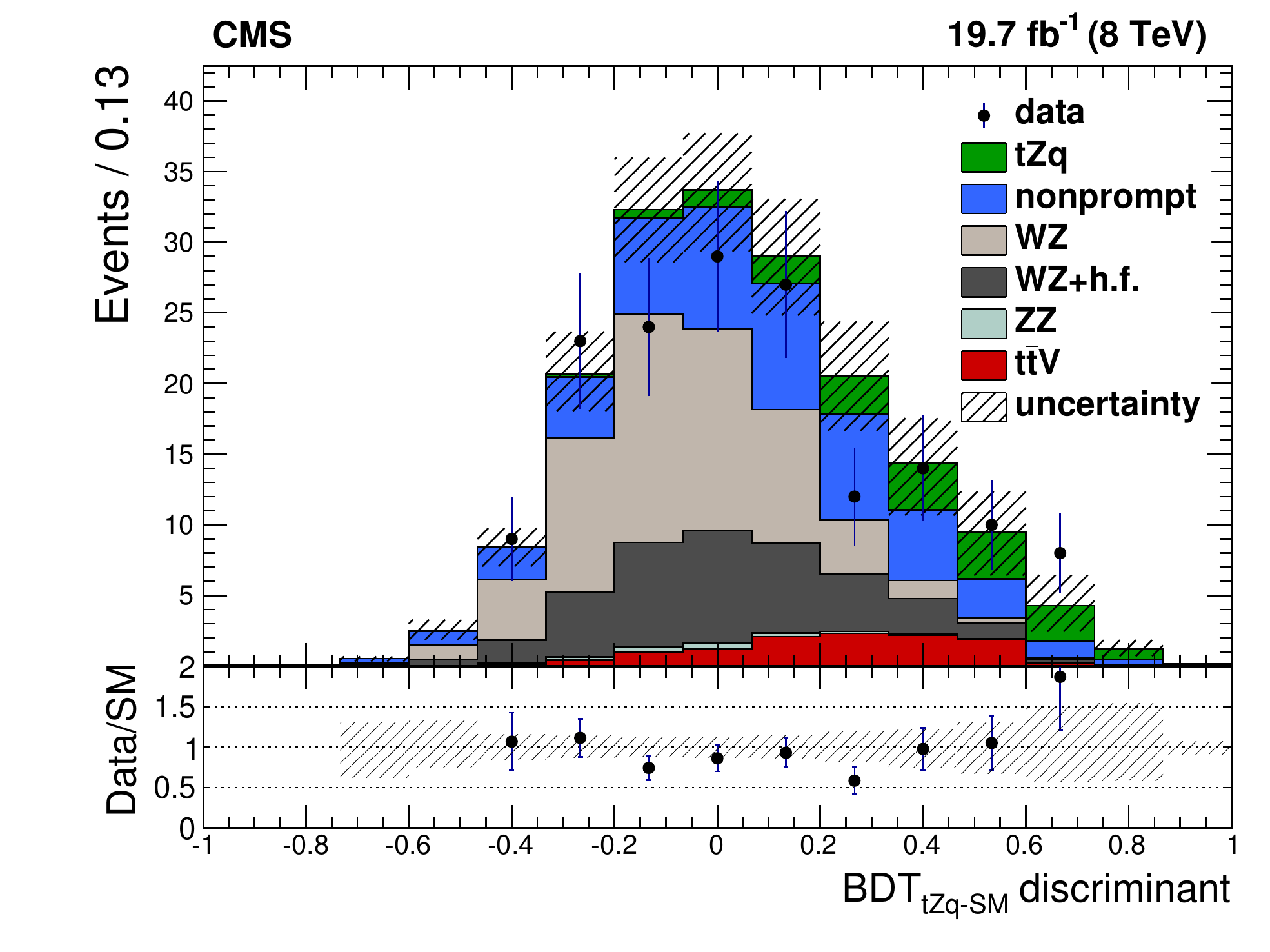}}
\end{center}
\caption{Data-to-prediction comparisons after performing the fit for \mwt distribution in the control region (left) and for the BDT$_{\mathrm{tZq}\text{-SM}}$ responses in the signal region (right). The four lepton channels are combined. The lower panels show the ratio between observed and predicted yields, including the total uncertainty on the prediction.}
\label{fig:post_tZqSM}
\end{figure}

\begin{figure*}[hbtp!]
\centering
{\includegraphics[width=0.45\linewidth]{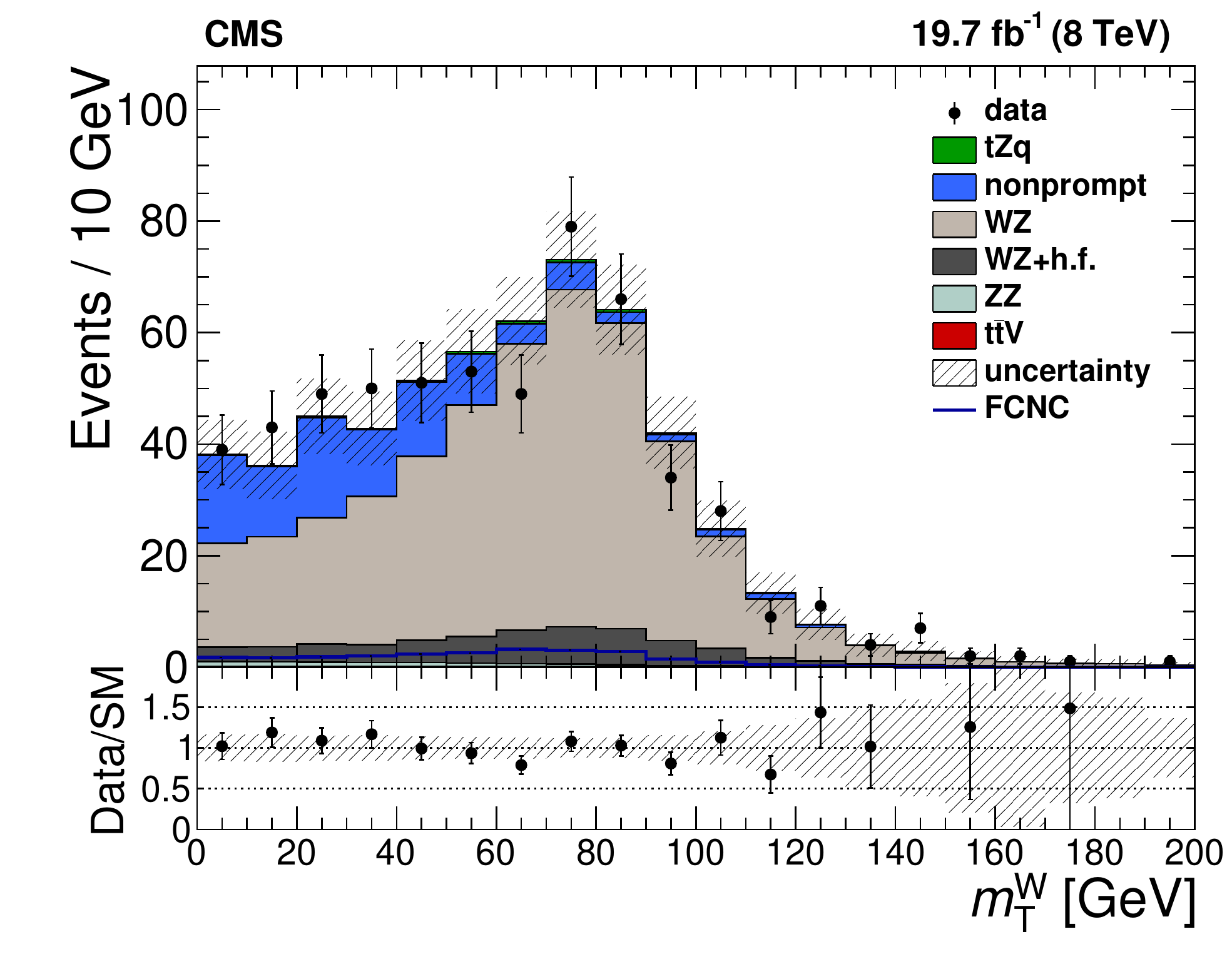}} 
{\includegraphics[width=0.45\linewidth]{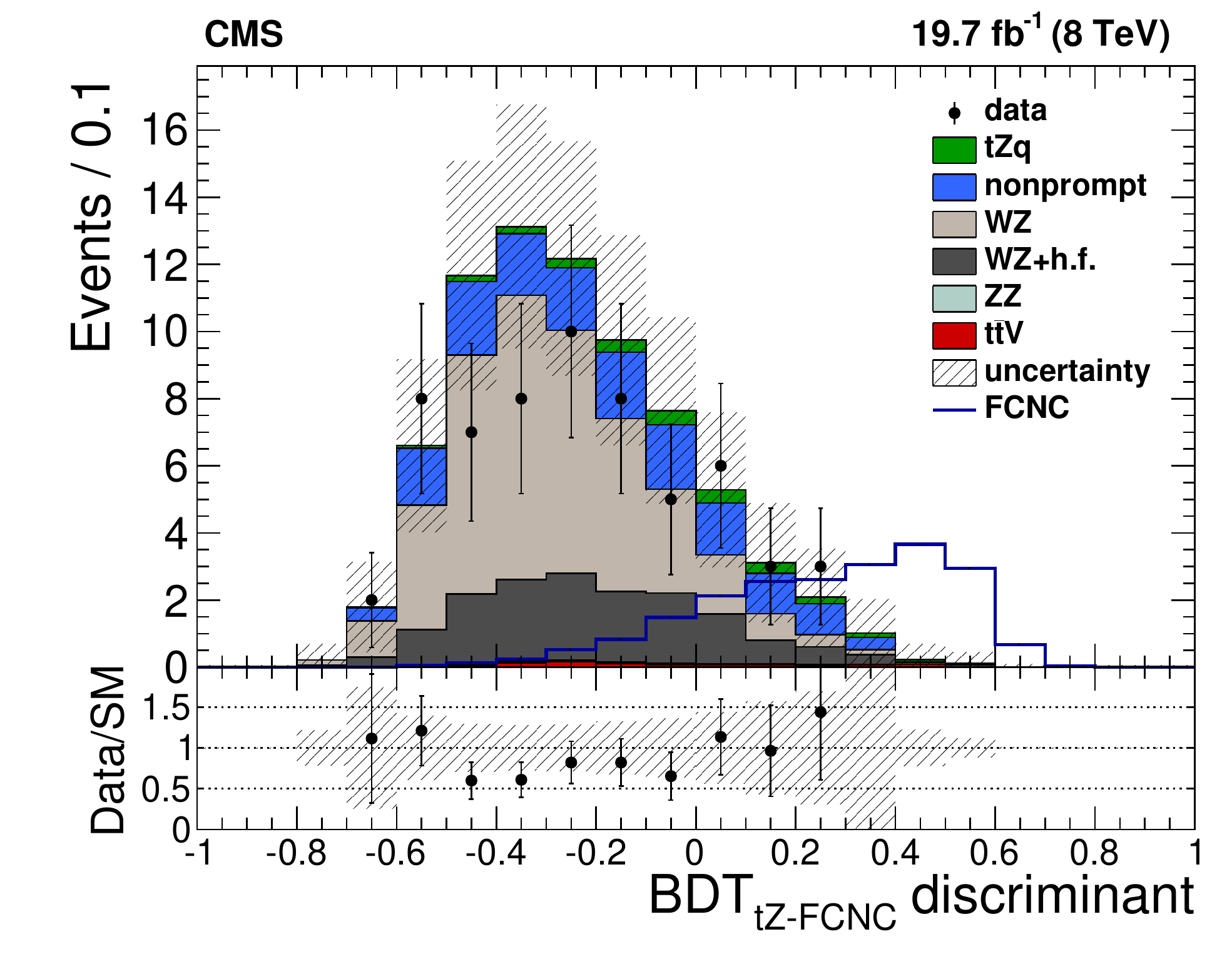}}\\
{\includegraphics[width=0.45\linewidth]{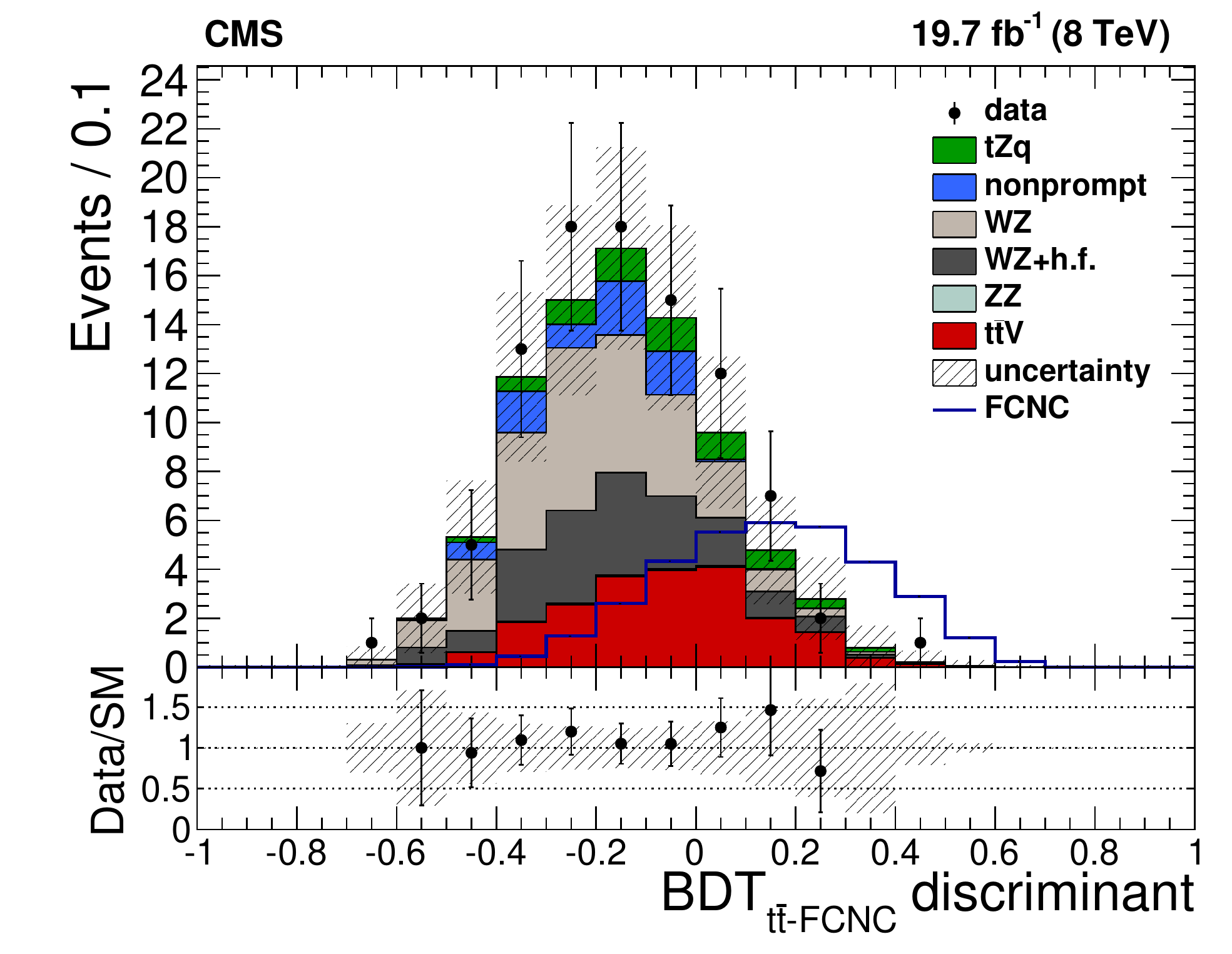}}\\
\caption{Data-to-prediction comparisons for the tZ-FCNC search after performing the fit for \mwt distribution in the control region (top-left), and for the BDT responses in the single top quark (BDT$_{\text{tZ-FCNC}}$) (top-right), and $ \ttbar $ (BDT$_{\text{\ttbar-FCNC}}$) (bottom), signal regions. An example of the predicted signal contribution for a value ${ \cal{B} }(\PQt \rightarrow \PZ \PQu) = 0.1\%$ (FCNC) is shown for illustration. The four channels are combined. The lower panels show the ratio between observed and predicted yields, including the total uncertainty on the prediction.}
\label{fig:postfit_plots}
\end{figure*}

In order to enhance the separation between signal and background processes, a multivariate discriminator is used in both the $ \PQt \PZ \PQq $-SM and FCNC searches. The discriminator is based on the BDT algorithm~\cite{BDT} implemented in the standard toolkit for multivariate analysis TMVA~\cite{Hocker:2007ht}. A range of different quantities are used as input variables for the BDTs. They are selected based on their discriminating power and include kinematic variables related to the top quark and the Z boson, such as \pt, pseudorapidity, and charge asymmetry $q_\ell \, |\eta|$, where $q$ and $\eta$ are the charge and $\eta$ of the lepton from the W decay, as well as jet properties, particularly those related to b tagging or the pseudorapidity of the recoiling jet. 
The BDTs are trained using half of the simulated samples for these processes and they are trained separately for each channel. The output discriminant distribution is then fitted, in the signal region, for each channel, to determine whether there are any signal events present in the data. The second half of the simulated samples are used to test that overtraining did not occur. 
For the SM search, the  BDT$_{\mathrm{tZq}\text{-SM}}$ is used to discriminate between the $ \PQt \PZ \PQq $-SM signal and the dominating $\ttbar \PZ$ and WZ background processes. The  BDT$_{\mathrm{tZq}\text{-SM}}$ distribution is fitted, together with the \mwt distribution in the control region. The results of the fits are presented in Fig.~\ref{fig:post_tZqSM} for the four channels combined. 
For the FCNC searches, the BDT$_{\text{tZ-FCNC}}$ and BDT$_{\text{\ttbar-FCNC}}$ are used to discriminate FCNC processes from the SM background processes. The  BDT$_{\text{tZ-FCNC}}$, and BDT$_{\text{\ttbar-FCNC}}$, distributions are fitted, together with the \mwt distribution in the control region. The results of the fits are presented in Fig.~\ref{fig:postfit_plots} for the four channels combined.
A number of different background processes are considered. These include \ttbar, single top quark, diboson, $ \ttbar\mathrm{V} $, and DY production. The contamination from W+jets events involves two nonprompt leptons and is found to be negligible. Diboson production is dominated by the WZ sample, which is split into two parts: the production of WZ events in association with light jets, or in association with heavy-flavour jets. The ZZ production contributes with a small number of background events. While the cross section of WW production is slightly higher than ZZ production, a nonprompt lepton would have to be selected to replicate the signal, making its contribution to the background negligible. The $ \ttbar $ SM and the DY backgrounds populate the signal region if they contain a reconstructed nonprompt lepton that passes the lepton identification and isolation selections; as the nonprompt lepton rates are not well modelled by the simulation, these backgrounds are estimated from data. The \mwt distribution is used as a discriminator in the background-enriched region to estimate the backgrounds related to nonprompt leptons, as well as the dominant WZ background. Both the shape and normalisation of the other backgrounds are estimated from simulation.

The normalisation of the nonprompt lepton and WZ background is estimated by fitting the \mwt distribution. The \mwt\ distribution peaks around the W mass for a lepton and \ptmiss\ from a W boson decay, while for nonprompt lepton backgrounds it peaks close to zero and falls rapidly. This difference in shape allows a simultaneous estimation of the nonprompt lepton and the WZ backgrounds to be made. In the \eemu\ and \mumue\ final states, the same-flavour opposite-sign leptons are assumed to come from the Z boson, hence the remaining lepton (third lepton) is assumed to come from the W boson and is used to compute the transverse mass. For the \eee\ and \mumumu\ final states, both opposite sign combinations are considered.
The normalised \mwt\ distributions (templates) for events containing a nonprompt lepton are obtained by inverting the isolation criteria on the third lepton. The resulting event sample is expected to be dominated by DY events, although a small number of $ \ttbar $ events are expected.
The signal is extracted by performing a simultaneous binned maximum-likelihood fit to the distributions of the signal samples and the background-enriched control region, using the two different discriminators. The background-enriched control region helps to constrain the backgrounds in the signal sample by means of nuisance parameters. A common fit is performed simultaneously for the four different final states (\eee, \eemu, \mumue, and \mumumu).
\begin{figure*}[hbtp!]
\centering
{\includegraphics[width=0.45\linewidth]{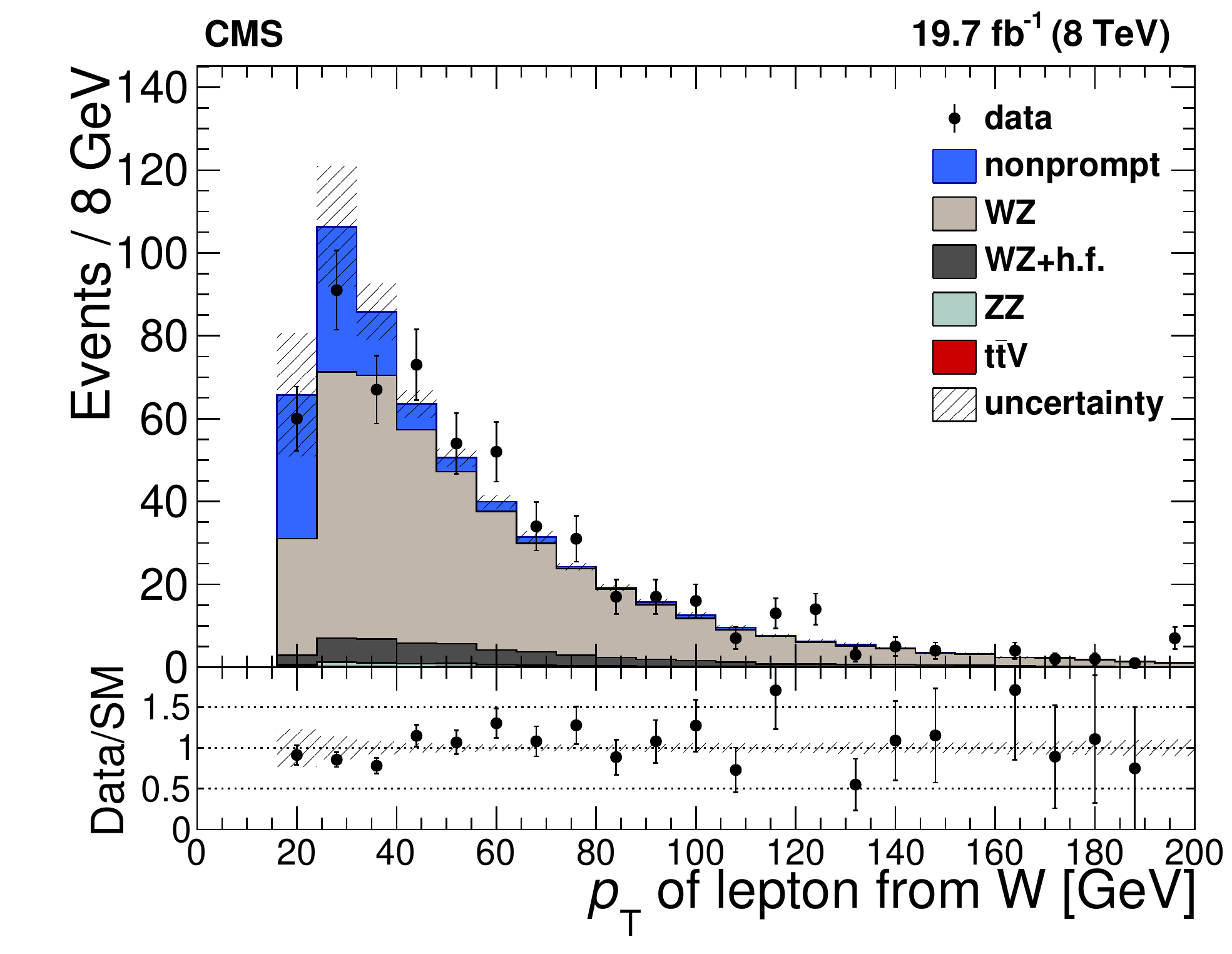}}    
{\includegraphics[width=0.45\linewidth]{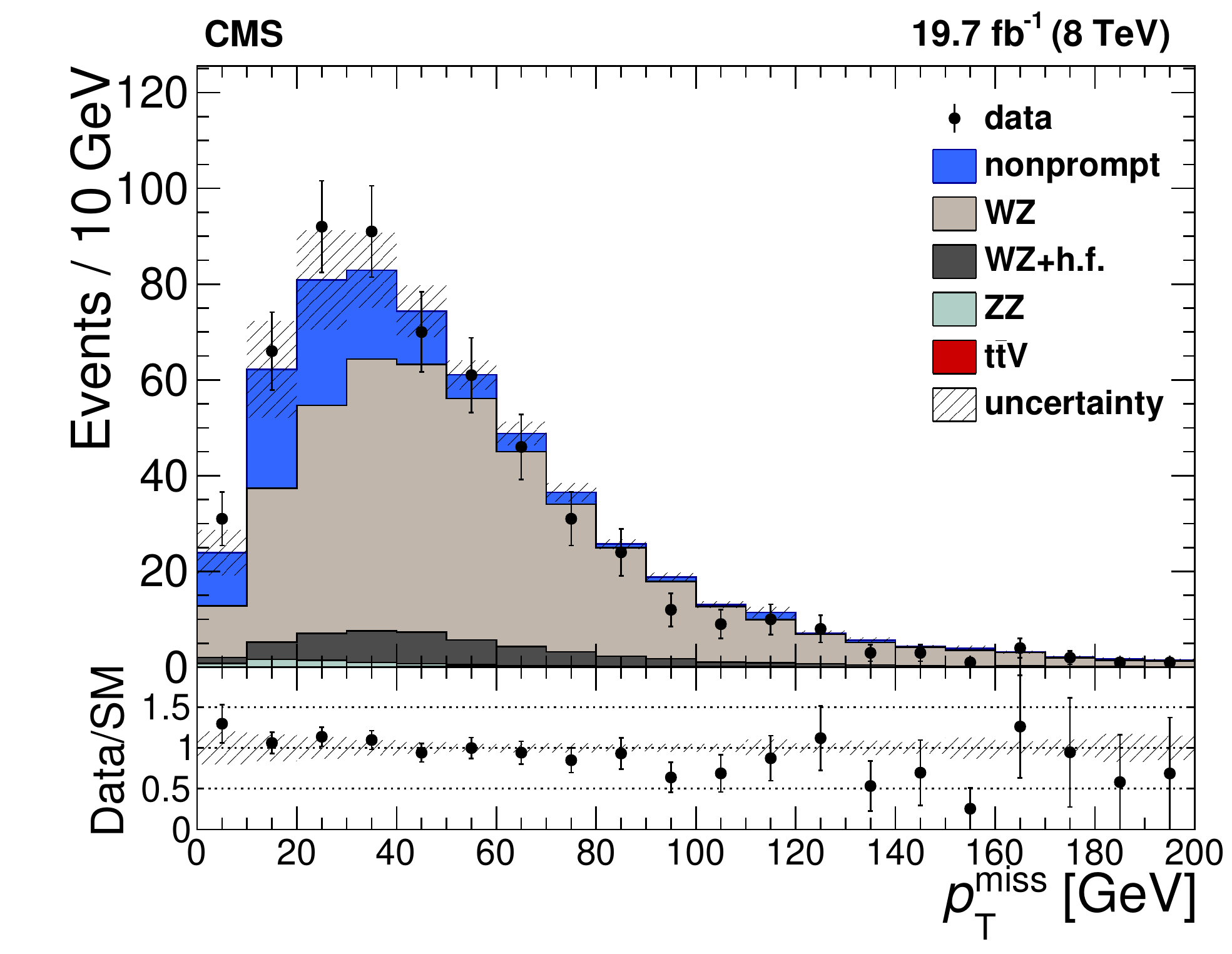}}\\
{\includegraphics[width=0.45\linewidth]{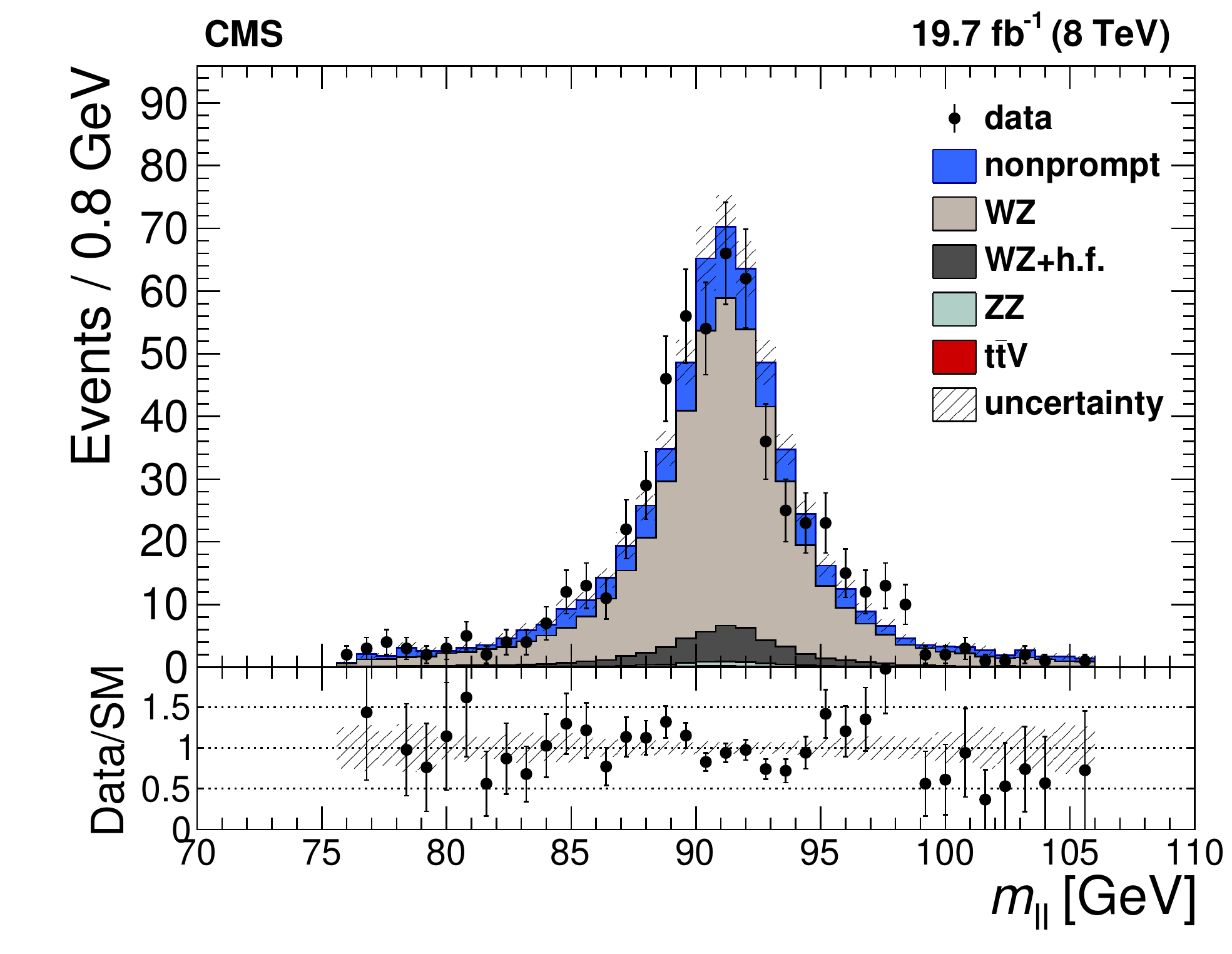}}\\
\caption{Data-to-prediction comparisons in the background-enriched samples, after  applying background normalisation scaling factors as described in the text, of the \pt of the lepton from the W boson (top-left), \ptmiss\ (top-right), and $m _{\ell \ell}$ (bottom). The four channels are combined. The lower panels show the ratio between observed and predicted yields, including the total uncertainty on the prediction. The distributions shown here are for the tZ-FCNC search, where WZ + h.f. denotes WZ + heavy flavour.}
\label{fig:mtwWZ}
\end{figure*} 
In order to validate the fit procedure, an additional fit is performed in the background-enriched region only and the background normalisations are extracted from this fit. These normalisations are used to compare the data to the predictions as shown in Fig.~\ref{fig:mtwWZ}. Reasonable agreement in normalisation and shape between data and predictions is found, validating the background model.
\section{Systematic uncertainties}
\label{sec:syst}
Different sources of systematic uncertainty are considered. They can affect the number of events passing the selection, the shape of the BDT response, or both.
\begin{itemize}
\item {\bf Luminosity measurement:}  The integrated luminosity measurement is extracted using the pixel cluster counting method~\cite{CMS-PAS-LUM-13-001}, with the corresponding uncertainty being ${\pm}2.6\%$.
\item {\bf Pileup estimation:}  The uncertainty in the average expected number of additional interactions per bunch crossing is ${\pm}5\%$.
\item {\bf Lepton trigger, reconstruction, and identification efficiency:}  To ensure that the efficiency of the dilepton triggers observed in data is properly reproduced, a set of data-to-simulation corrections is applied to all simulated events ; likewise, an additional set of corrections (\pt- and $\eta$- dependent) is used to ensure that the efficiency for reconstructing and identifying leptons observed in the data is correctly reproduced in the simulation.  The corrections are varied by their corresponding uncertainties, which amounts to about 4\% per event for the trigger selection and 2\% per event for the lepton selection.  For the tZ-FCNC production the trigger selection is extended, which increases the acceptance and in turn leads to a reduction in the trigger uncertainty.
\item {\bf Jet energy scale (JES), jet energy resolution (JER), and missing transverse momentum:}  In all simulated events, all the reconstructed jet four-momenta
are simultaneously varied by the uncertainties associated with the jet energy scale and resolution.  Changing the jet momenta in this fashion causes a corresponding change in the total momentum in the transverse plane, thus affecting \ptmiss\ as well.  The contribution to \ptmiss\ that is not from particles identified as leptons or photons, or that are not clustered into jets is varied by ${\pm}10\%$~\cite{Chatrchyan:2011ds}.
\item {\bf b tagging:}  The b tagging and misidentification efficiencies are estimated using control samples~\cite{Chatrchyan:2012jua}.  The resulting corrections are applied to all simulated samples to ensure that they reproduce the efficiencies in data.  The corrections are varied by ${\pm}1$ standard deviation ($\sigma$).
\item {\bf Background normalisation:}  The normalisation of the nonprompt lepton and WZ background processes are estimated from data while performing the final fit. The normalisation uncertainties in the backgrounds estimated from simulation are taken as 30\%. The WZ + jets sample is split into two parts: WZ + light-flavour jets and WZ + heavy-flavour (b and c) jets. The normalisations of these two backgrounds, which are treated separately, are left free in the fit.
\item {\rm {\bf Z boson \boldmath{ \pt} \bf :}}   Uncertainty coming from the Z boson \pt reweighting is accounted for by not applying, or applying twice, the reweighting.
\item {\bf Physics process modelling:}  The renormalisation and factorisation scales used in the WZ, $ \PQt \PZ \PQq $-SM and tZ-FCNC signal simulation, as well as for the  $\ttbar\PZ$ simulated samples, are multiplied or divided by a factor of two, and the corresponding variations are considered as shape systematic uncertainties. The procedure used in {\sc pythia} to match the partons in the matrix-element calculation with those in the parton showering includes a number of scale thresholds.  These are varied in the simulated WZ sample and the resulting variation is taken as the associated systematic uncertainty.
\item {\bf PDFs:}  The nominal PDF sets used for the analyses described in this paper are quoted in Section~\ref{sec:pdf}. In order to compute the corresponding uncertainty, simulated events are reweighted by using the eigenvalues associated to each PDF set. The corresponding variations are summed in quadrature and the results are compared with the nominal prediction. Uncertainties estimated from different PDF sets are also compared and the largest uncertainty is taken.
\item {\bf Simulated sample size:}  The statistical uncertainty arising from the limited size of the simulated samples is taken as a source of systematic uncertainty using the "Barlow-Beeston light" method \cite{BarlowBeeston}.  
\end{itemize}
The systematic sources, variation and type (shape/normalisation) are summarised in Table~\ref{tab:systcorr}. For a given source of systematic uncertainty there is 100\% correlation between the 4 channels, except for the lepton misidentification where the \mumumu\ and \eemu\ channels are 100\% correlated and the \mumue\ and \eee\ channels are 100\% correlated, due to the isolation inversion of the lepton candidate from the W decay. 
\begin{table}[!ht]
\begin{center}
\topcaption{The systematic sources, variation and type, which represent how the uncertainty is treated in the likelihood fit.} 
\begin{tabular}{l|c|c}
\hline
Systematic source & Variation & Type \\
\hline
Z+jets, $ \ttbar $            & ${\pm}30\%$ & norm.\\
Muon misidentification     & floating in the fit & norm.\\
Electron misidentification & floating in the fit & norm.\\
Z \pt                      & ${\pm} 1\sigma$ & shape\\
WZ+l jets norm.            & floating in the fit & norm.\\
WZ+l jets matching         & ${\pm} 1\sigma$ & shape\\
WZ+l jets scale            & Q$^2\!\times\!4$, Q$^2$/4 & shape\\
WZ+hf jets norm.           & floating in the fit & norm.\\
WZ+hf jets matching        & ${\pm} 1 \sigma $& shape\\
WZ+hf jets scale           & Q$^2\!\times\!4$, Q$^2$/4 & shape\\
$ \PQt \PZ \PQq $                        & ${\pm}30\%$& norm.\\
$ \PQt \PZ \PQq $ scale                  & Q$^2\!\times\!4$, Q$^2$/4 & norm.+shape\\
ZZ                         & ${\pm}30\%$ & norm. \\
Single top                 & ${\pm}30\%$ & norm. \\
$ \ttbar\mathrm{V} $                 & ${\pm}30\%$ & norm. \\
\hline
Trigger                  & ${\pm} 1 \sigma$  & norm.\\
Lepton selection               & $\pm$1\% & norm.+shape\\
JES                      & ${\pm} 1 \sigma(\pt, \eta)$  & norm.+shape\\
JER                      & ${\pm} 1 \sigma(\pt, \eta)$  &  norm.+shape\\ 
Uncertainty \ptmiss             & ${\pm}10\%$  &  norm.+shape\\ 
b tagging                & ${\pm} 1 \sigma(\pt, \eta)$ & norm.+shape\\ 
Pileup                   & ${\pm} 1 \sigma$ & norm.+shape\\
PDF                      & ${\pm} 1\sigma$ & norm.+shape\\          
\hline
& & \\ [-2.4ex]
tZ-FCNC scale             & Q$^2\!\times\!4$, Q$^2$/4 & norm.+shape\\ 
\hline
Luminosity                     & ${\pm}2.6\%$    & norm.\\
\hline
\end{tabular}
\label{tab:systcorr}
\end{center}
\end{table}
\section{Results}
\label{sec:results}
The fit is performed on the BDT discriminant distributions in the signal samples, and on the \mwt\ distributions in the background-enriched sample, for each of the four final states (\eee, \eemu, \mumue, and \mumumu). This is implemented using the Theta program~\cite{theta}, with most of the systematic uncertainties treated as nuisance parameters.
Prior to fitting, the templates for each background process are scaled to correspond to the predicted SM cross section, including all relevant corrections, and the integrated luminosity of the data sample used for the analysis. The systematic uncertainties discussed in Section~\ref{sec:syst} are included in the fit. For each source of systematic uncertainty, $u$, a nuisance parameter, $\theta_{u}$, is introduced. Systematic uncertainties can affect the rate of events and/or the shape of the template distribution. The data are used to constrain the nuisance parameters for all systematic uncertainties except for those related to the physics process modelling and PDF parameters. The significance is calculated using a Bayesian technique.
\subsection{Search for \texorpdfstring{$ \PQt \PZ \PQq $}{tZq}-SM production}

\label{sec:tZqshapeAna}
By performing a simultaneous fit on the \mwt distribution in the background-enriched sample and on the BDT outputs in the signal region, the number of events in excess of the background-only hypothesis is determined. This excess can then be compared to the SM expectation for $ \PQt \PZ \PQq $ production in order to measure the cross section. The efficiency times acceptance for the BDT-based analysis is 0.10 for the inclusive cross section.
\begin{figure}[htbp!]
\begin{center}
{\includegraphics[width=0.45\linewidth]{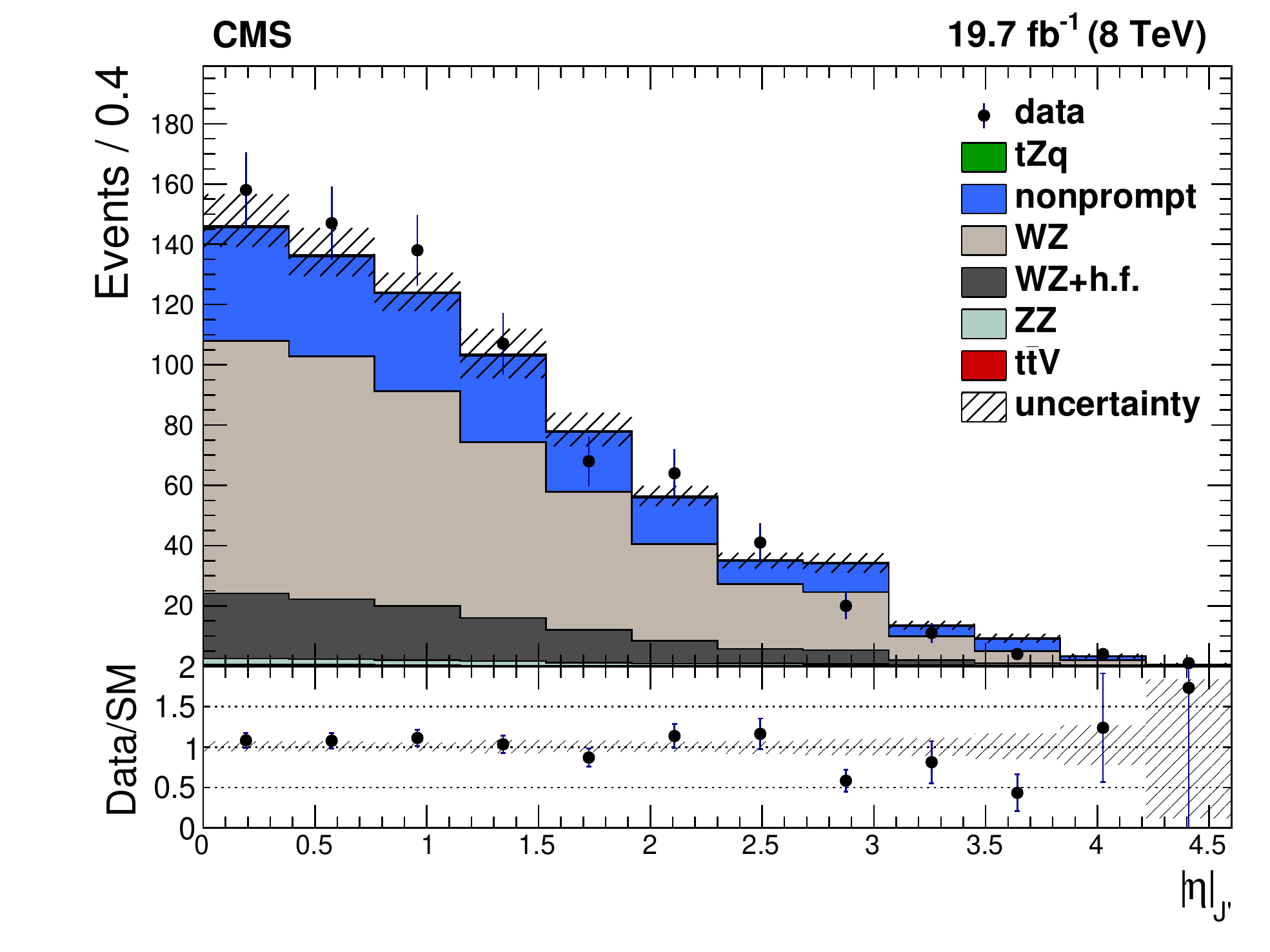}}
{\includegraphics[width=0.45\linewidth]{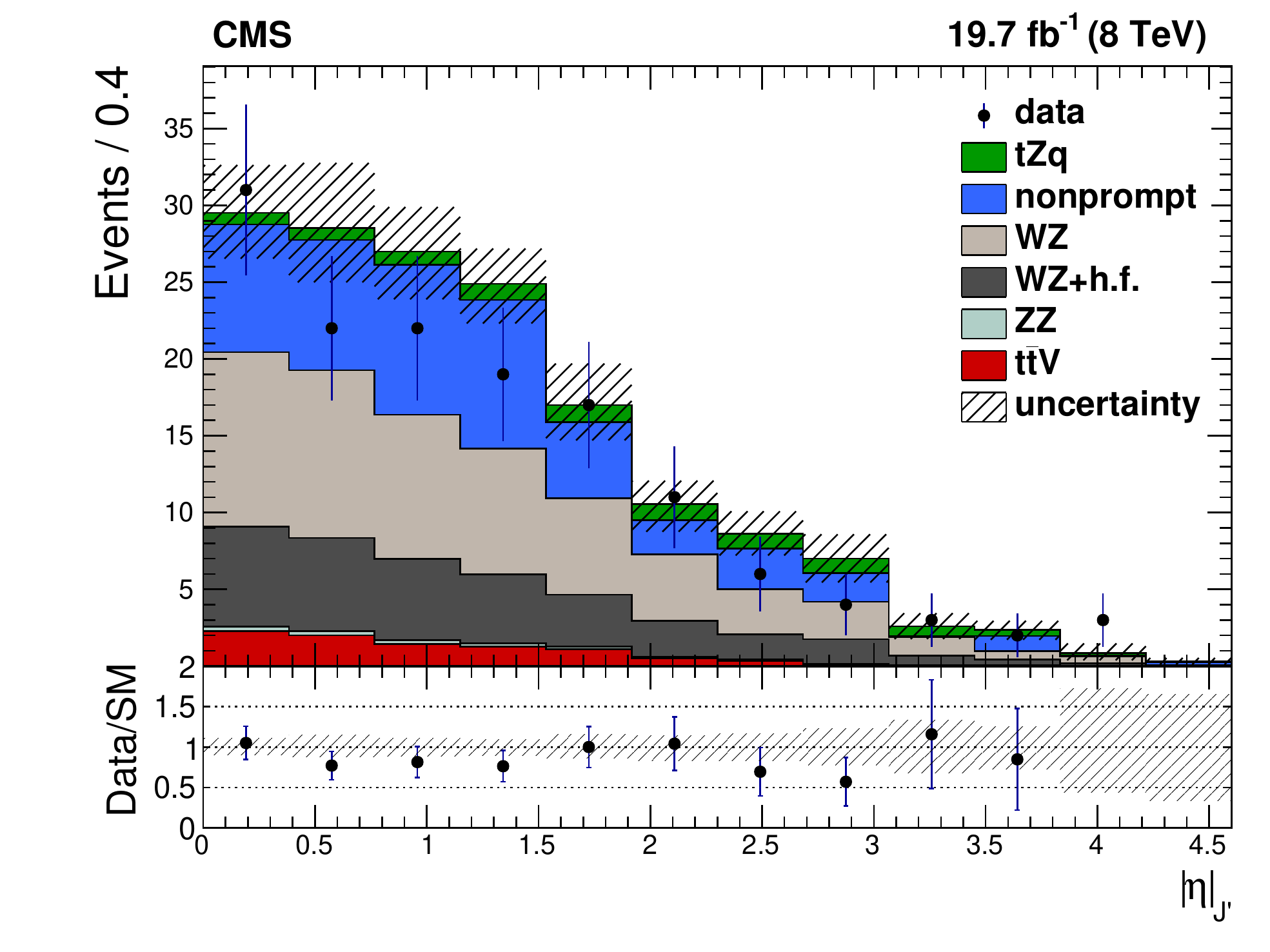}}
\end{center}
\caption{Data-to-prediction comparisons after performing the fit for the $|\eta|$ distribution of the recoiling jet in the control region (left), and the signal region (right). The four lepton channels are combined. The lower panels show the ratio between observed and predicted yields, including the total uncertainty on the prediction.}
\label{fig:post_jet}
\end{figure}
The measured cross sections for the individual channels and the channels combined are shown in Table~\ref{tab:shapetxs}. The combined measured signal $ \PQt \PZ \PQq $ cross section is found to be $ 10^{+8}_{-7}$ fb and is consistent with the SM prediction of 8.2\unit{fb} with a theoretical uncertainty of less than 10\%. For illustration, the data-to-prediction comparisons, including the post-fit uncertainties, are presented in Fig.~\ref{fig:post_jet} for the $|\eta|$ distribution of the leading jet not originating from the top quark decay ($\eta_{\rm J '}$) in the control region and in the signal region.
The corresponding observed and expected significances are $2.4$ and $1.8$ standard deviations, respectively, with the expected significance having a one standard deviation range of [0.4 -- 2.7] at 68\%~CL. 
The observed signal exclusion limit on the $ \PQt \PZ \PQq $ cross section is 21\unit{fb} at 95\%~CL.
\begin{table}[!h]
\renewcommand{\arraystretch}{1.2}
\centering
\topcaption{The measured cross sections, together with their total uncertainties, for the individual channels and the channels combined for the BDT-based analysis.}
\begin{tabular}{c|c}
Channel           & Cross section (fb)     \\ 
\hline 
$\Pe\Pe\Pe$               &   $0^{+9}$ \\ 
$\Pe\Pe\mu$           & $11^{+13}_{-10}$  \\ 
$\mu\mu\Pe$         &  $24^{+19}_{-16}$  \\ 
$\mu \mu \mu$     &   $5^{+9}_{-5}$  \\  
\hline
Combined fit & $10^{+8}_{-7}$  \\ 
\end{tabular}
\label{tab:shapetxs}
\end{table}
As a cross-check, the search for $ \PQt \PZ \PQq $-SM is also performed using a counting experiment.  The main differences in the event selection compared to the BDT-based analysis are a tighter electron isolation requirement, $I_{ \rm rel} < 0.1$, and a tighter $m_{\ell \ell}$ selection $ 78  < m_{\ell \ell} <  102\GeV$. For this analysis, the WZ background is estimated by counting the number of events in a region enriched in WZ events, defined by inverting the b tagging requirements. Contamination of other sub-dominant processes is subtracted using the prediction of the simulation and a systematic uncertainty is estimated by varying their yields according to their respective uncertainties. Additional systematic uncertainties due to the WZ modelling are accounted for by considering renormalisation and factorisation scale variations as well as matching threshold variations. 
For the cross-check analysis the total expected number of events is $15.4 \pm 0.5$, dominated by $\ttbar\PZ$ events ($5.2 \pm 0.3$)  and WZ events ($3.6 \pm 0.2$). The contribution from ZZ, \ttbar, and DY events is $2.7 \pm 0.3$, and the contribution from $ \ttbar\PW $ events is $0.5 \pm 0.02$. The expected number of signal events is $3.4 \pm 0.1$. A total of 20 events passing all signal selections are observed in the data. The efficiency times acceptance for the counting experiment is 0.021 for the inclusive cross section. The measured cross sections for each channel, and the combination of channels, is calculated using the \textsc{RooStats} package~\cite{RooStats}. The results obtained  are shown in Table~\ref{tab:cutcountxs}.
\begin{table}[!h]
\renewcommand{\arraystretch}{1.2}
\centering
\topcaption{The measured cross sections for the individual channels and the channels combined for the counting analysis.}
\begin{tabular}{c|c}
Channel       & Cross section (fb) \\ 
\hline
eee            & $\rm 29^{+32}_{-24} (stat)^{+8}_{-7}\ (syst)$  \\ 
ee$\mu$            & $\rm \phantom{0} 6^{+23}_{-6} (stat)^{+4}_{-3}\ (syst)$  \\ 
$\mu \mu$e            & $\rm \ 19^{+24}_{-18} (stat) {\pm} 5 (syst)$  \\ 
$\mu \mu \mu$            & $\rm 20^{+19}_{-15} (stat)^{+4}_{-3}\ (syst)$   \\ 
\hline
Combined fit    & $\rm \ 18^{+11}_{-9} (stat) {\pm} 4 (syst)$  \\ 
\end{tabular}
\label{tab:cutcountxs}
\end{table}
The cross section is measured to be $\rm 18^{+11}_{-9} (stat) \pm 4 (syst)$ fb, in agreement with the SM prediction and with the BDT-based result. The corresponding signal significance is observed to be 1.8 standard deviations, while the expected significance is 0.8 standard deviations, with a 68\% CL range of [0 --1.59].
\subsection{Search for \texorpdfstring{tZ-FCNC}~ production}
\begin{figure}[htbp!]
\begin{center}
{\includegraphics[width=0.45\linewidth]{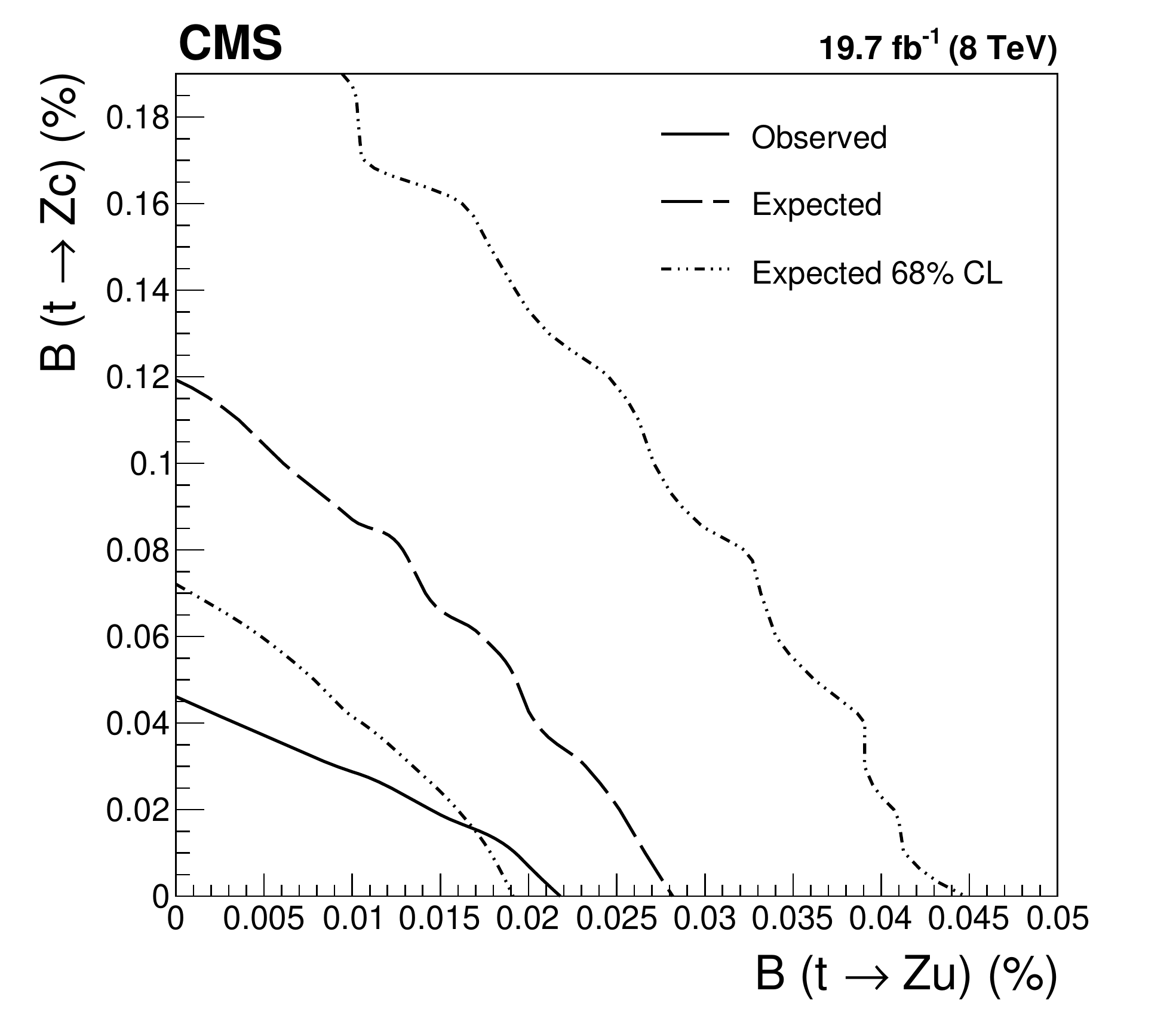}}
\end{center}
\caption{The expected and observed exclusion limits at 95\% CL on ${ \cal{B} } (\PQt \rightarrow \PZ \PQc)$ as a function of the limits on $ {\cal{B}}( \PQt \rightarrow \PZ \PQu)$. The expected 68\% CL is also shown. }
\label{fig:2Dlimits}
\end{figure}

To search for tZ-FCNC interactions, the single-top-quark-FCNC, \ttbar-FCNC and background-enriched samples are combined in a single fit. The result of the fit is consistent with the SM-only hypothesis. 
Exclusion limits at 95\% CL for tZ-FCNC are calculated by performing simultaneously the fit in the single-top-quark-FCNC-, \ttbar-FCNC-, and WZ-enriched regions.
The limits are calculated for different combinations of $ \PQt \PZ \PQu $ and $ \PQt \PZ \PQc $ anomalous couplings, as shown in Fig.~\ref{fig:2Dlimits}.
\begin{table}[htbp!]
\renewcommand{\arraystretch}{1.2}
\begin{center}
\topcaption{Expected and observed 95\% exclusion limits on the branching fraction of the tZ-FCNC couplings.} 
\label{tab:limits_FCNC}
\begin{tabular}{c|c|c|c|c}
Branching fraction & Expected & 68\% CL range & 95\% CL range & Observed\\ 
\hline
${ \cal{B} } ( \PQt \rightarrow \PZ \PQu) $ (\%)	    &	0.027	&  0.018 -- 0.042    & 0.014 -- 0.065  &  0.022  \\ 
${ \cal{B} } ( \PQt \rightarrow \PZ \PQc) $ (\%)	    &	0.118	&  0.071 -- 0.222    & 0.049 -- 0.484  &  0.049 \\
\end{tabular}
\end{center}
\end{table}
The independent exclusion limits are summarised in Table~\ref{tab:limits_FCNC} where the branching fraction of the coupling not under consideration is assumed to be zero. A more stringent limit is observed on the $ \PQt \PZ \PQu $ couplings compared to the $ \PQt \PZ \PQc $ couplings as a result of the larger cross section for tZ-FCNC in the $ \PQt \PZ \PQu $ channel. The limits are ${ \cal{B} } ( \PQt \rightarrow \PZ \PQu)  < 0.022\%$ and ${ \cal{B} } ( \PQt \rightarrow \PZ \PQc)  < 0.049\%$, which improve the previous limits set by the CMS Collaboration~\cite{CMSTOPFCNC} by about a factor of two.
\section{Summary}
A search for the associated production of a top quark and a Z boson, as predicted by the standard model was performed with the full CMS data set collected at 8\TeV, corresponding to an integrated luminosity of 19.7\fbinv. An events yield compatible with $ \PQt \PZ \PQq $ standard model production is observed, and the corresponding cross section is measured to be $10^{+8}_{-7}\unit{fb}$. The corresponding observed and expected significances are 2.4 and 1.8 standard deviations, respectively.
A search for tZ production produced via flavour-changing neutral current interactions, either in single-top-quark or $\ttbar$ production modes, was also performed. For this search the standard model $ \PQt \PZ \PQq $ process was considered as a background. No evidence for tZ-FCNC interactions is found, and limits at 95\% confidence level are set on the branching fraction for the decay of a top quark into a Z boson and a quark. The limits are ${ \cal{B} } ( \PQt \rightarrow \PZ \PQu) < 0.022\%$ and ${ \cal{B} } ( \PQt \rightarrow \PZ \PQc)  < 0.049\%$, which improve the previous limits set by the CMS Collaboration by about a factor of two.
\begin{acknowledgments}
\hyphenation{Bundes-ministerium Forschungs-gemeinschaft Forschungs-zentren Rachada-pisek} We congratulate our colleagues in the CERN accelerator departments for the excellent performance of the LHC and thank the technical and administrative staffs at CERN and at other CMS institutes for their contributions to the success of the CMS effort. In addition, we gratefully acknowledge the computing centres and personnel of the Worldwide LHC Computing Grid for delivering so effectively the computing infrastructure essential to our analyses. Finally, we acknowledge the enduring support for the construction and operation of the LHC and the CMS detector provided by the following funding agencies: the Austrian Federal Ministry of Science, Research and Economy and the Austrian Science Fund; the Belgian Fonds de la Recherche Scientifique, and Fonds voor Wetenschappelijk Onderzoek; the Brazilian Funding Agencies (CNPq, CAPES, FAPERJ, and FAPESP); the Bulgarian Ministry of Education and Science; CERN; the Chinese Academy of Sciences, Ministry of Science and Technology, and National Natural Science Foundation of China; the Colombian Funding Agency (COLCIENCIAS); the Croatian Ministry of Science, Education and Sport, and the Croatian Science Foundation; the Research Promotion Foundation, Cyprus; the Secretariat for Higher Education, Science, Technology and Innovation, Ecuador; the Ministry of Education and Research, Estonian Research Council via IUT23-4 and IUT23-6 and European Regional Development Fund, Estonia; the Academy of Finland, Finnish Ministry of Education and Culture, and Helsinki Institute of Physics; the Institut National de Physique Nucl\'eaire et de Physique des Particules~/~CNRS, and Commissariat \`a l'\'Energie Atomique et aux \'Energies Alternatives~/~CEA, France; the Bundesministerium f\"ur Bildung und Forschung, Deutsche Forschungsgemeinschaft, and Helmholtz-Gemeinschaft Deutscher Forschungszentren, Germany; the General Secretariat for Research and Technology, Greece; the National Scientific Research Foundation, and National Innovation Office, Hungary; the Department of Atomic Energy and the Department of Science and Technology, India; the Institute for Studies in Theoretical Physics and Mathematics, Iran; the Science Foundation, Ireland; the Istituto Nazionale di Fisica Nucleare, Italy; the Ministry of Science, ICT and Future Planning, and National Research Foundation (NRF), Republic of Korea; the Lithuanian Academy of Sciences; the Ministry of Education, and University of Malaya (Malaysia); the Mexican Funding Agencies (BUAP, CINVESTAV, CONACYT, LNS, SEP, and UASLP-FAI); the Ministry of Business, Innovation and Employment, New Zealand; the Pakistan Atomic Energy Commission; the Ministry of Science and Higher Education and the National Science Centre, Poland; the Funda\c{c}\~ao para a Ci\^encia e a Tecnologia, Portugal; JINR, Dubna; the Ministry of Education and Science of the Russian Federation, the Federal Agency of Atomic Energy of the Russian Federation, Russian Academy of Sciences, the Russian Foundation for Basic Research and the Russian Competitiveness Program of NRNU ?MEPhI?; the Ministry of Education, Science and Technological Development of Serbia; the Secretar\'{\i}a de Estado de Investigaci\'on, Desarrollo e Innovaci\'on, Programa Consolider-Ingenio 2010, Plan de Ciencia, Tecnología e Innovaci\'on 2013-2017 del Principado de Asturias and Fondo Europeo de Desarrollo Regional, Spain; the Swiss Funding Agencies (ETH Board, ETH Zurich, PSI, SNF, UniZH, Canton Zurich, and SER); the Ministry of Science and Technology, Taipei; the Thailand Center of Excellence in Physics, the Institute for the Promotion of Teaching Science and Technology of Thailand, Special Task Force for Activating Research and the National Science and Technology Development Agency of Thailand; the Scientific and Technical Research Council of Turkey, and Turkish Atomic Energy Authority; the National Academy of Sciences of Ukraine, and State Fund for Fundamental Researches, Ukraine; the Science and Technology Facilities Council, UK; the US Department of Energy, and the US National Science Foundation.

Individuals have received support from the Marie-Curie programme and the European Research Council and EPLANET (European Union); the Leventis Foundation; the A. P. Sloan Foundation; the Alexander von Humboldt Foundation; the Belgian Federal Science Policy Office; the Fonds pour la Formation \`a la Recherche dans l'Industrie et dans l'Agriculture (FRIA-Belgium); the Agentschap voor Innovatie door Wetenschap en Technologie (IWT-Belgium); the Ministry of Education, Youth and Sports (MEYS) of the Czech Republic; the Council of Science and Industrial Research, India; the HOMING PLUS programme of the Foundation for Polish Science, cofinanced from European Union, Regional Development Fund, the Mobility Plus programme of the Ministry of Science and Higher Education, the National Science Center (Poland), contracts Harmonia 2014/14/M/ST2/00428, Opus 2014/13/B/ST2/02543, 2014/15/B/ST2/03998, and 2015/19/B/ST2/02861, Sonata-bis 2012/07/E/ST2/01406; the National Priorities Research Program by Qatar National Research Fund; the Programa Clar\'in-COFUND del Principado de Asturias; the Thalis and Aristeia programmes cofinanced by EU-ESF and the Greek NSRF; the Rachadapisek Sompot Fund for Postdoctoral Fellowship, Chulalongkorn University and the Chulalongkorn Academic into Its 2nd Century Project Advancement Project (Thailand); and the Welch Foundation, contract C-1845.

\end{acknowledgments}
\clearpage

\bibliography{auto_generated}
\cleardoublepage \appendix\section{The CMS Collaboration \label{app:collab}}\begin{sloppypar}\hyphenpenalty=5000\widowpenalty=500\clubpenalty=5000\textbf{Yerevan Physics Institute,  Yerevan,  Armenia}\\*[0pt]
A.M.~Sirunyan, A.~Tumasyan
\vskip\cmsinstskip
\textbf{Institut f\"{u}r Hochenergiephysik,  Wien,  Austria}\\*[0pt]
W.~Adam, E.~Asilar, T.~Bergauer, J.~Brandstetter, E.~Brondolin, M.~Dragicevic, J.~Er\"{o}, M.~Flechl, M.~Friedl, R.~Fr\"{u}hwirth\cmsAuthorMark{1}, V.M.~Ghete, C.~Hartl, N.~H\"{o}rmann, J.~Hrubec, M.~Jeitler\cmsAuthorMark{1}, A.~K\"{o}nig, I.~Kr\"{a}tschmer, D.~Liko, T.~Matsushita, I.~Mikulec, D.~Rabady, N.~Rad, B.~Rahbaran, H.~Rohringer, J.~Schieck\cmsAuthorMark{1}, J.~Strauss, W.~Waltenberger, C.-E.~Wulz\cmsAuthorMark{1}
% ...
\vskip\cmsinstskip
% ...
1:~~Also at Vienna University of Technology, Vienna, Austria\\
% ...

\end{sloppypar}
\end{document}